\documentclass{ieeeaccess}
\usepackage{cite}
\usepackage{amsmath,amssymb,amsfonts}
\usepackage{algorithmic}
\usepackage{graphicx}
\usepackage{array,multirow,graphicx}
\usepackage{float}
\usepackage{xspace}
\usepackage{blindtext}
\usepackage{hyperref}
\usepackage{textcomp}
\usepackage{amsmath}
\usepackage{pdfpages}
\usepackage{float, subfig}
\usepackage{hyperref}
\def\BibTeX{{\rm B\kern-.05em{\sc i\kern-.025em b}\kern-.08em
    T\kern-.1667em\lower.7ex\hbox{E}\kern-.125emX}}
\begin{document}
\history{Date of publication xxxx 00, 0000, date of current version xxxx 00, 0000.}
\doi{10.1109/ACCESS.2017.DOI}

\title{An Accurate and Explainable Deep Learning System Improves Interobserver Agreement in the Interpretation of Chest Radiograph}
\author{\uppercase{Hieu H. Pham}\authorrefmark{1,2}, \IEEEmembership{Member, IEEE},
\uppercase{Ha Q. Nguyen\authorrefmark{3,4}},
\uppercase{Hieu T. Nguyen}\authorrefmark{3,4},
\uppercase{Linh T. Le}\authorrefmark{5},
\uppercase{Lam Khanh}\authorrefmark{6},
}

\address[1]{College of Engineering and Computer Science, VinUniversity, Hanoi, Vietnam}
\address[2]{VinUni-Illinois Smart Health Center, VinUniversity, Hanoi, Vietnam}
\address[3]{Smart Health Center, VinBigData JSC, Hanoi, Vietnam}
\address[4]{Vingroup Big Data Institute, Hanoi, Vietnam}
\address[5]{Hanoi Medical University Hospital, Department of Radiology, Hanoi, Vietnam}
\address[6]{Hospital 108, Department of Radiology, Hanoi, Vietnam}

\tfootnote{This research was supported by Vingroup Big Data Institute (VinBigData).}

\markboth
{Author \headeretal: Preparation of Papers for IEEE TRANSACTIONS and JOURNALS}
{Author \headeretal: Preparation of Papers for IEEE TRANSACTIONS and JOURNALS}
\corresp{Corresponding author: Hieu H. Pham (e-mail: hieu.ph@vinuni.edu.vn).}

\begin{abstract}
Interpretation of chest radiographs (CXR) is a difficult but essential task for detecting thoracic abnormalities. Recent artificial intelligence (AI) algorithms have achieved radiologist-level performance on various medical classification tasks. However, only a few studies addressed the localization of abnormal findings from CXR scans, which is essential in explaining the image-level classification to radiologists. Additionally, the actual impact of AI algorithms on the diagnostic performance of radiologists in clinical practice remains relatively unclear. To bridge these gaps, we developed an explainable deep learning system called VinDr-CXR that can classify a CXR scan into multiple thoracic diseases and, at the same time, localize most types of critical findings on the image. VinDr-CXR was trained on 51,485 CXR scans with radiologist-provided bounding box annotations. It demonstrated a comparable performance to experienced radiologists in classifying 6 common thoracic diseases on a retrospective validation set of 3,000 CXR scans, with a mean area under the receiver operating characteristic curve (AUROC) of 0.967 (95\% confidence interval [CI]: 0.958--0.975). The VinDr-CXR was also externally validated in independent patient cohorts and showed its robustness. For the localization task with 14 types of lesions, our free-response receiver operating characteristic (FROC) analysis showed that the VinDr-CXR achieved a sensitivity of 80.2\% at the rate of 1.0 false-positive lesion identified per scan. A prospective study was also conducted to measure the clinical impact of the VinDr-CXR in assisting six experienced radiologists. The results indicated that the proposed system, when used as a diagnosis supporting  tool, significantly improved the agreement between radiologists themselves with an increase of 1.5\% in mean Fleiss' Kappa. We also observed that, after the radiologists consulted VinDr-CXR's suggestions, the agreement between each of them and the system was remarkably increased by 3.3\% in mean Cohen's Kappa. Altogether, our results highlight the potentials of the proposed deep learning system as an effective  assistant to radiologists in clinical practice. Part of the dataset used for developing the VinDr-CXR system has been made publicly available at \href{https://physionet.org/content/vindr-cxr/1.0.0/}{https://physionet.org/content/vindr-cxr/1.0.0/}.
\end{abstract}

\titlepgskip=-15pt
\maketitle
\section{Introduction}
\label{sec:introduction}
\PARstart{C}{ommon} chest pathologies affect several hundred million people worldwide and kill several million cases every 
year~\cite{TB,Cancer}. They are the leading cause of death and impose an immense worldwide health burden. Diagnosis of thoracic diseases is a crucial clinical task for physicians. Currently, chest X-ray (CXR) radiography is the primary imaging modality used for screening, triaging, and diagnosing varieties of lung conditions~\cite{corne2015chest} such as pneumothorax, pneumonia, tuberculosis (TB), pleural effusion, atelectasis, emphysema, and cancers, etc. However, the CXR interpretation is a complicated task, which requires an in-depth understanding of radiologic signs in thoracic imaging~\cite{delrue2011difficulties,fitzgerald2001error,manning2004detection,carmody1980analysis}. A previous study~\cite{donald2012common} reported that 22\% of all errors in diagnostic radiology were made in the CXR interpretation. A recent work~\cite{gavelli2000sensitivity} showed that 19\%--26\% of lung cancers visible on CXR images were missed at the first reading. Furthermore, interpreting CXR scans usually is highly dependent on the observer and has a poor interagreement between physicians~\cite{peng2017does}. The interobserver agreement was considered poor to moderate depending on the type of findings~\cite{moncada2011reading}; this rate can be lower in local hospitals, leading to unfavorable consequences.

Advanced machine learning algorithms have recently shown their significant potential in medical image analysis~\cite{alghamdi2021deep,litjens2017survey,pham2021interpreting,allaouzi2019novel,nguyen2021vindr_spinexr,arias2020artificial,tran2021learning,kwon2021pggan,tran2022transparency,nguyen2021vindr_ribCXR,ahishali2021advance,le2022learning,rahman2020reliable,dao2022phase,nguyen2022deployment,sakib2020dl,nguyen2021novel} due to the availability of large-scale datasets~\cite{wang2017chestx,bustos2020padchest,irvin2019chexpert,johnson2019mimic,nguyen2020vindrcxr,nguyen2022vindr_pcxr,nguyen2022vindr_mammo,nguyen2021vindr_spinexr,pham2021dicom}. In particular, previous studies~\cite{rajpurkar2017chexnet,rajpurkar2018deep,irvin2019chexpert,majkowska2020chest,rajpurkar2020chexpedition,tang2020automated,ting2017development,zago2018retinal,liu2020deep,esteva2017dermatologist} have indicated that a deep learning (DL) system trained on a large-scale, annotated medical imaging dataset can reach a level of performance comparable to practicing radiologists in detecting common thoracic diseases~\cite{rajpurkar2017chexnet,rajpurkar2018deep,irvin2019chexpert,majkowska2020chest,rajpurkar2020chexpedition,tang2020automated}, analyzing retinal images~\cite{ting2017development,zago2018retinal}, or diagnosing skin cancers~\cite{liu2020deep,esteva2017dermatologist}. 

However, the actual impact of DL systems in clinical practice remains unclear, and large-scale clinical evaluations of these such systems are limited. Hence, despite many promising results and increasing performances that have been published, very few DL algorithms have reached clinical implementation. We observe that multiple factors slow or impede artificial intelligence (AI) transfer into clinical practice. First, the development of an accurate and robust DL system requires a large number of annotated CXR scans from diverse sources. The creation of large-scale, high-quality datasets of annotated images is costly and challenging. Meanwhile, public datasets are limited, and their labels are unreliable since they are produced by automated rule-based labelers~\cite{wang2017chestx,irvin2019chexpert,smit2020chexbert}. Previous evidences~\cite{rayner,oakden2020exploring,majkowska2020chest} showed that training DL systems on small datasets and low-quality annotations raises concerns about the robustness of those systems in real clinical contexts. Second, few clinical evaluations of DL-assisted diagnostic algorithms have been performed, and most of them are not even prospective and at high risk of bias~\cite{nagendran2020artificial}. Third, although DL systems can outperform physicians in specific clinical tasks, the lack of explanatory power~\cite{amann2020explainability,holzinger2019causability} became a key obstacle to convince medical experts, who must understand how and why DL models have made a prediction. We believe that an accurate assessment of CXRs requires both detection of abnormal findings and a correct decision at the disease level. Hence, the provision of accurate and interpretable visualizations of lung abnormalities is a crucial step towards the clinical translation of DL systems.

To address these gaps, we introduce in this study a fully automated DL system, namely VinDr-CXR, for chest radiograph interpretation. The VinDr-CXR is designed to simultaneously classify six common lung diseases and localize 14 important findings from CXR scans. The development and evaluation of VinDr-CXR are based on large-scale medical imaging analysis and state-of-the-art DL algorithms. Specifically, we use a patient dataset from multiple hospitals in Vietnam, containing 51,485 manually annotated CXR studies, to train the DL system. To evaluate the performance of the proposed system, we compare the model's performance with that of human experts in a benchmark study using the consensus annotations provided by 5 experienced radiologists on a retrospective dataset as the reference standard. To validate the generalization capability of VinDr-CXR, we compute its performance on various external datasets. The results confirm that our framework is accurate and robust across multiple populations and settings. To demonstrate the clinical value of VinDr-CXR as an assistant to radiologists, we conduct a prospective study at two hospitals in Vietnam and investigate the inter-rater agreement on CXR interpretation with and without VinDr-CXR's assistance. We further calculate the change in the agreement rate between VinDr-CXR itself and each radiologist before and after he/she consults the system's suggestions. Experiments indicate that the proposed DL system provides meaningful supports for radiologists in detecting thoracic diseases in a real-world clinical setting.

\section{Materials and methods}

Our study was approved by the Institutional Review Boards (IRB) of the Hanoi Medical University Hospital (HMUH) and Hospital 108 (H108). In addition, the requirement for obtaining informed patient consent was waived due to the observational nature of this study. Under these approvals, raw CXR images in the Digital Imaging and Communications in Medicine (DICOM) format were collected retrospectively. Protected health information (PHI) has been de-identified to comply with the regulations of the U.S. HIPAA~\cite{assistance2003summary}, European GDPR~\cite{gdpr}, and the local privacy laws~\cite{article_8}. In this section, we describe the methodology for developing and validating the VinDr-CXR system. First, we provide an overview of the proposed approach. Next, we describe details of datasets used in this study. Then, the development of DL algorithms is introduced. Last, we describe the experimental design for our reader study and statistical analysis methods used for model evaluation.

\subsection{Overview of approach}

\begin{figure*}
  \centering
  \includegraphics[width=17cm,height=15cm]{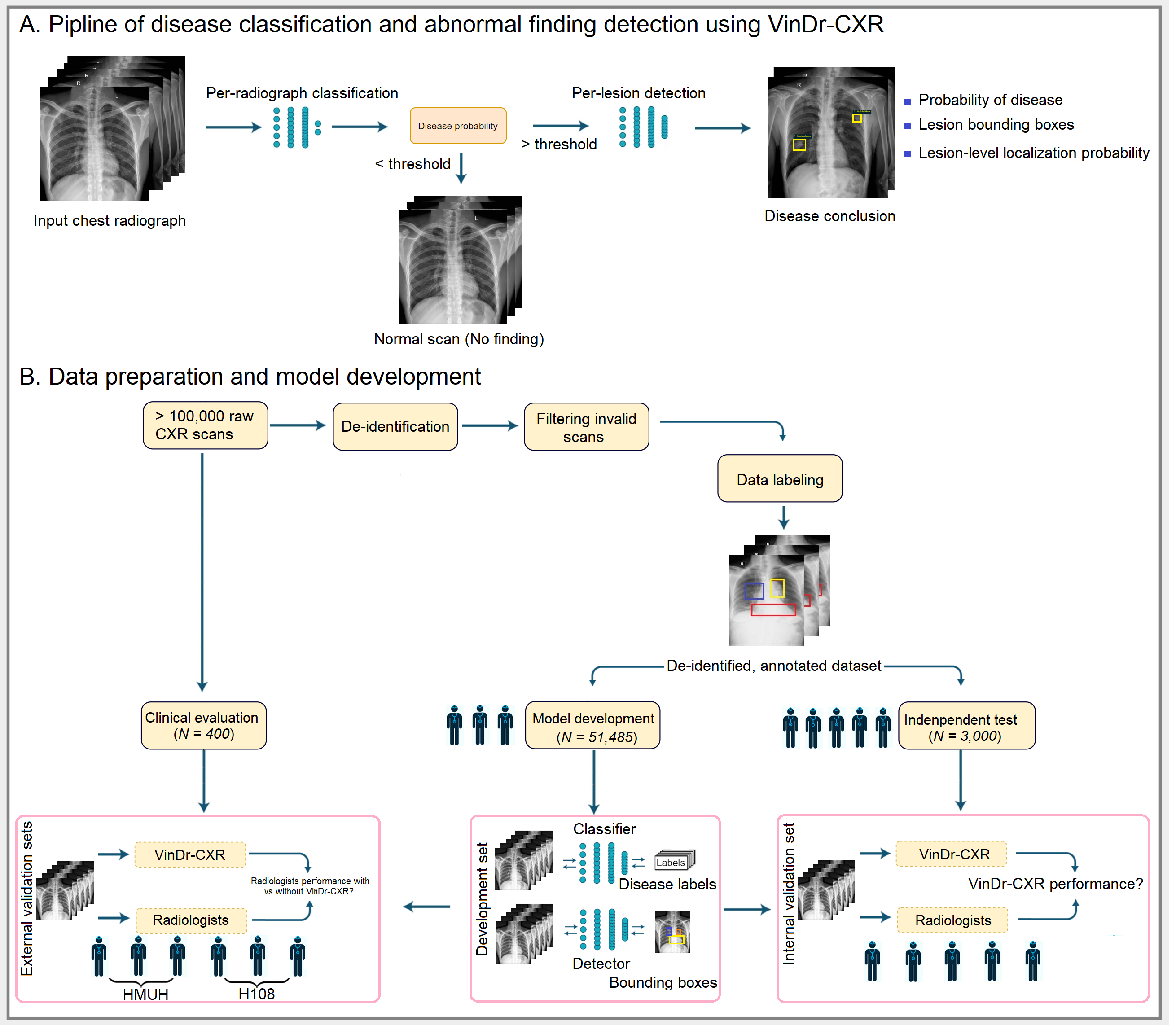}
 \caption{\normalsize{Overview of the proposed approach. (A) The pipeline of the VinDr-CXR framework for lung disease classification and localization using CXR images. A classifier takes as input one CXR scan and predicts its probability of abnormality. A detector takes an abnormal scan as input and provides bounding boxes with probabilities of abnormal findings at the lesion-level. The final prediction of the whole framework comes up with a complete description consisting of a disease conclusion, and abnormal findings. (B) Summary of the VinDr-CXR development and evaluation. The raw images in the DICOM standard were collected retrospectively from the picture archiving and communication system (PACS) of the HMUH and H108 hospitals. The raw images were then de-identified and all out-of-distribution samples were removed. An in-house web-based labeling tool called VinDr Lab was designed to annotate imaging data. In the labeling process, each CXR scan from the development set was labeled by a group of three experienced radiologists for the presence of 22 preliminary findings and six disease impressions. Meanwhile, CXR scans from the internal validation set were annotated by a consensus of five experienced radiologists. Finally, 51,485 CXR images were used to develop DL algorithms, and 3,000 studies were used for validation. External evaluations were also performed on CheXpert~\cite{irvin2019chexpert} (\textit{N = 200}) and CheXphoto~\cite{phillips2020chexphoto} (\textit{N = 200}) datasets to evaluate the robustness of the VinDr-CXR across multiple hospitals. A reader study was designed and performed at the HMUH and H108 (\textit{N = 400}) to investigate the impact of the VinDr-CXR in clinical practice.}}
  \label{fig:study_overview}
\end{figure*}

We present in this section VinDr-CXR, a DL-based framework for the per-radiograph classification of common lung diseases and the detection of abnormal findings on CXRs. This framework includes two major components. First, an image-level classification network accepts a CXR scan as input and predicts whether it could be normal or abnormal. Second, a lesion-level detection network receives an abnormal CXR scan as input from the classifier and provides the location of abnormal findings via bounding box predictions. An overview of the proposed approach is illustrated in Figure~\ref{fig:study_overview}A. The core of the VinDr-CXR system is based on state-of-the-art DL networks for image classification and object detection tasks, named EfficientNet~\cite{tan2019efficientnet} and EfficientDet~\cite{tan2020efficientdet}, respectively. The classification network is trained with image-level labels to distinguish six common lung diseases, including \textit{Pneumonia}, \textit{Tuberculosis}, \textit{Lung Tumor}, \textit{Pleural Effusion}, \textit{Other Diseases}, and \textit{No Finding}. Meanwhile, the detection network is trained with lesion-level annotations to localize 14 critical findings from the CXR images, i.e., \textit{Cardiomegaly}, \textit{Opacity}, \textit{Consolidation}, \textit{Atelectasis}, \textit{Pneumothorax}, \textit{Pleural Effusion}, \textit{Aortic Enlargement}, \textit{Interstitial Lung Disease (ILD)}, \textit{Infiltration}, \textit{Nodule/Mass}, \textit{Pulmonary Fibrosis}, \textit{Pleural Thickening}, \textit{Calcification}, and \textit{Other Lesions}. To develop DL algorithms, a total of 51,485 anteroposterior (AP) and posteroanterior (PA) CXR scans have been retrospectively collected from the HMUH and H108, which then manually annotated by expert radiologists. To evaluate the diagnostic accuracy of the VinDr-CXR, we compare its prediction with ground truth on an internal validation set of 3,000 studies, which was separate from the training set. Additionally, external datasets including CheXpert (\textit{N} = 200) CheXphoto (\textit{N} = 200) are used to verify the robustness of the VinDr-CXR for cross-site validation. Finally, we investigate the actual impact of the VinDr-CXR on clinical practice through a large reader study (\textit{N} = 400). The inter-rater agreement among radiologists as well as the rate of agreement between VinDr-CXR and radiologists are then be assessed. Figure~\ref{fig:study_overview}B shows an overview of the development and evaluation of the VinDr-CXR framework.

\subsection*{Datasets for VinDr-CXR development and validation}

\subsubsection*{VinDr-CXR dataset}

To develop and validate the VinDr-CXR, we retrospectively collected a total of more than 100,000 anterior–posterior (AP) and posterior–anterior (PA) CXR scans of adult patients (aged $>$ 10 years). The imaging data were in the DICOM format and performed at two major hospitals (\textit{i.e.}, the HMUH and H108) between January 1, 2018 and December 31, 2020. In addition, CXR studies were acquired from a wide diversity of scanners and manufacturers such as Phillips, GE, Fujifilm, Siemens, Toshiba, Canon, and Samsung. Out-of-distribution samples (e.g., images with poor quality or invalid) were excluded via a manual inspection process. A group of 17 expert radiologists has labeled a portion of the raw dataset, and each has at least ten years of experience. Overall, the dataset contains 54,485 CXR studies (mean age 43.77 years; 47.79\% female patients) that meet the study criteria.  A development set consists of 51,485 studies used to optimize DL algorithms and an internal validation set of 3,000 studies to evaluate models' performance. All CXRs were performed in independent patients for the validation set and did not overlap with the development set to avoid bias. Each CXR scan in the development set was labeled by three independent radiologists, while a panel of five board-certified radiologists labeled each case in the internal validation set, and their consensus established the reference standard. Ground truth was established at the image-level for the classification task and pixel-level for the localization task. Our dataset was labeled for the presence of 28 labels. We defined for the first time two different types of labels in CXRs: (1) global labels that are image-level labels representing diseases or impressions and (2) local labels (lesion-level annotations) that are critical findings or lesions that occur in CXRs. The participating radiologists provide precisely the location of lesions or abnormalities via bounding box annotations for local labels. The data collection and annotation process is summarized in Figure~\ref{fig:study_overview}B. The cohort demographic information and statistics of development and internal validation data sets are summarized in Table~\ref{data-characs}. Full details of the dataset collection and labeling process can be found in our previous study~\cite{nguyen2020vindrcxr}.

\subsubsection*{External datasets}
We used two public CXR datasets to assess the accuracy and efficiency of the VinDr-CXR system across populations. Our external validation tests used data from patients in the CheXpert~\cite{irvin2019chexpert} (\textit{N = 200}) and CheXphoto~\cite{phillips2020chexphoto} (\textit{N = 200}) datasets. These datasets shared several disease labels with the VinDr-CXR dataset, such as Pleural Effusion, Pneumonia, and No Finding. The CheXpert~\cite{irvin2019chexpert} is a large public dataset for the CXR interpretation performed between 2002 and 2017 at Stanford Hospital, USA. The CheXpert validation set contains 200 studies, for which the ground-truth label of each study is obtained by taking the majority vote of three board-certified radiologists. The CheXphoto~\cite{phillips2020chexphoto} is a recently published CXR dataset for the automated interpretation of photos of CXR through cell phone photography. The CheXphoto validation set comprises natural photos of all 200 studies in the CheXpert validation set. It can be used as a resource for testing the robustness of DL algorithms on smartphone photos of CXRs. Additional details of the external datasets are provided in Irvin \textit{et al}.~\cite{irvin2019chexpert} and Phillips \textit{et al}.~\cite{phillips2020chexphoto}.

\begin{table*}
\centering
\textsf{\scriptsize{
\caption{Characteristics of the datasets used for VinDr-CXR development and internal validation.}
\label{data-characs}
\setlength{\tabcolsep}{3pt}
\begin{tabular}{p{15pt} | p{150pt}|p{150pt}|p{150pt}}
\hline
&  \textbf{Characteristics} & 
 \textbf{Development set} & 
 \textbf{Internal validation set}  \\
\hline
  \parbox[t]{2mm}{\multirow{12}{*}{\rotatebox[origin=c]{90}{\textbf{Collection statistics}}}} &  &  & \\
  &   Years &   2018 to 2020 &   2018 to 2020\\
&   Number of studies &  51,485 &  3,000 \\
&   Number of images &   51,941 &   3,000\\
&  Number of abnormal studies &  26,887 &  949 \\
&   Number of normal studies &  24,598 &   2,051\\
&   Number of annotators per scan  &  3 &   5  \\
&   Image size (pixel$\times$pixel, mean) &   2,729 $\times$ 2,395 &  2,748 $\times$ 2,394 \\
&   Age (years, mean)*  &   51.28 &  31.80  \\
&   Male (\%)* &   52.77  &   55.90    \\
&   Female (\%)* &   47.23  &  44.10  \\
&   Data size (GiB) &    509.5 &    31.3 \\
&  &  &  \\
\hline
\parbox[t]{3mm}{\multirow{22.5}{*}{\rotatebox[origin=c]{90}{\textbf{Local labels}}}}  &  &   &  \\
&   1. Aortic Enlargement                &  9,183 (17.68\%) &  221 (07.37\%) \\
&   2. Atelectasis                       &  1,346 (2.59\%) &  96  (3.20\%) \\
&   3. Cardiomegaly                      &  8,359 (16.09\%) &  310 (10.33\%) \\
&   4. Calcification                     &  3,154 (6.07\%)  &  232 (7.73\%) \\
&   5. Clavicle Fracture                 &  300   (0.58\%)  &  2   (0.07\%) \\
&   6. Consolidation                     &  2,125 (4.09\%)  &  126 (4.20\%) \\
&   7. Edema                             &  61    (0.12\%)  &  0   (0\%) \\
&   8. Emphysema                         &  1,007 (1.94\%)  &  4   (0.13\%) \\
&   9. Enlarged PA                       &  451   (0.87\%)  &  9   (0.30\%) \\
&  10. Interstitial Lung Disease (ILD)   &  5,946 (11.45\%) &  316 (10.53\%) \\
&  11. Infiltration                      &  3,254 (6.26\%)  &  79  (2.63\%) \\
&  12. Lung Cavity                       &  175   (0.34\%)  &  10  (0.33\%) \\
&  13. Lung Cyst                         &  129   (0.25\%)  &  3   (0.10\%) \\
&  14. Lung Opacity                      &  4,680 (9.01\%)  &  95  (3.17\%) \\
&  15. Mediastinal Shift                 &  706   (1.36\%)  &  20  (0.67\%) \\
&  16. Nodule/Mass                       &  5,521 (10.63\%) &  286 (9.53\%) \\
&  17. Pulmonary Fibrosis                &  6,919 (13.32\%) &  358 (11.93\%) \\
&  18. Pneumothorax                      &  321   (0.62\%)  &  25  (0.83\%) \\
&  19. Pleural Thickening                &  7,899 (15.21\%) &  240 (8.00\%) \\
&  20. Pleural Effusion                  &  4,554 (8.76\%)  &  136 (4.53\%) \\
&  21. Rib Fracture                      &  1,241 (2.39\%)  &  17  (0.57\%) \\
&  22. Other Lesions                     &  3,851 (7.41\%)  &  112 (3.73\%) \\
&  &  & 
\\
\hline 
\parbox[t]{2mm}{\multirow{8}{*}{\rotatebox[origin=c]{90}{\textbf{Global labels}}}} &  &   &  \\
&  23. Lung  Tumor       &  1,650  (3.18\%) &  80    (2.67\%) \\
&  24. Pneumonia         &  3,827  (7.37\%)  &  246   (8.20\%) \\
&  25. Tuberculosis      &  2,130  (4.10\%)  &  164   (5.47\%) \\
&  26. Other Diseases    &  18,848 (36.29\%) &  657   (21.90\%) \\
&  27. COPD              &  388    (0.75\%)  &  2     (0.07\%)    \\
&  28. No Finding        &  16,461 (31.7\%)  &  2,051 (68.37\%) \\
&         &  &   \\
\hline
\end{tabular}}}
\begin{flushleft} \scriptsize{(*) The calculations were performed based on the number of CXR scans for which sex and age were known. For global labels, the number of positive examples was reported based on the majority vote of binarized labels provided by radiologists. For local labels, the percentage rate of number bounding boxes on the number of studies was reported. Several abnormalities (Clavicle Fracture, Edema, Emphysema, Enlarged PA, Lung Cavity, Lung Cyst, Mediastinal Shift, Rib Fracture, and COPD) were not considered during model training because the number of positive examples in the internal validation set was very limited. The “No Finding” label was intended to capture the absence of all findings and pathologies.} \end{flushleft}
\end{table*}

\subsection*{Model development and training}

To develop the VinDr-CXR system, we used a total of 51,485 annotated CXR scans from the development set. The system takes a CXR as input, and outputs are both disease classification and lesion localization. The whole VinDr-CXR architecture consists of two subnetworks, including a classification network and a detection network (see Figure~\ref{fig:study_overview}A). We trained the EfficientNet-B6~\cite{tan2019efficientnet} model on the CXR images with a size of 1024$\times$1024 pixels to classify common chest diseases. The EfficientNet~\cite{tan2019efficientnet} was well-known as a state-of-the-art DL architecture for image recognition tasks. It can achieve a high level of accuracy while requiring less computational cost for model training. We used mean binary cross-entropy loss to optimize the network in a supervised manner using image-level annotations. To localize abnormal findings, we deployed EfficientDet-D6~\cite{tan2020efficientdet}, a recent advance of DL-based detector for the object detection tasks. The per-lesion annotations provided by radiologists were used to optimize the EfficientDet-D6~\cite{tan2020efficientdet} network. To reduce the impact of class imbalance, we adopted the focal loss~\cite{lin2017focal} to optimize the detection network's weights. Several data augmentation strategies have been applied to minimize the risk of over-fitting in both two networks. Both the classification and detection networks were implemented using Python 3.7 (\url{https://www.python.org/}), PyTorch 1.6 (\url{https://pytorch.org/}), and trained on an NVIDIA V100 32GB GPU. A detailed description of the model development and training is provided in the Appendices.

\subsection*{Reader study}

To validate the effectiveness of the proposed DL approach, we conducted a reader study to assess the actual impact of the VinDr-CXR on the agreement of participating radiologists. We describe the reader study as bellows.\\[0.3cm]
\setlength\parindent{0pt}
\textbf{\textit{Data collection}}. For clinical evaluation, 400 CXR examinations were collected retrospectively from the HMUH and H108 under IRB approvals. These examinations were acquired between March 2021 and June 2021 after the training process of the VinDr-CXR has been completed. Among 400 CXR studies, half (\textit{N = 200}) was obtained from the HMUH, and the rest (\textit{N = 200}) was from the H108. Data sampling was conducted based on the actual distributions at the hospitals. This ensures that the imaging data are representative of the real-world conditions in which the DL algorithms will be deployed. In addition, the CXR scans will be used to evaluate the agreement among participating readers. Hence, we did not establish a reference standard for the collected data.

\textbf{\textit{Reader selection}.} We recruited a group of six board-certified radiologists from the radiology departments of the HMUH and H108 to participate in our observer performance test. All participating radiologists were trained in CXR interpretation and had an average of 15.5 years of clinical experience interpreting thoracic diseases (range 10–22 years). In addition, the readers read an average of 25,000 CXR scans each year (range 15,000–40,000). Table~\ref{rad-characs} shows the characteristics of radiologists who participated in our reader study.

\begin{table}
\centering
\scriptsize{
\caption{Characteristics of the participating radiologists. Mean annual diagnostic volumes were estimated based on the number of CXR scans interpreting.}
\label{rad-characs}
\setlength{\tabcolsep}{3pt}
\begin{tabular}{|p{65pt}|p{54pt}|p{104pt}|}
\hline
 \textbf{Reader} &  \textbf{Year's experience} &   \textbf{Annual diagnostic volume (studies)} \\
\hline
 Radiologist 1 (H108) &   \hspace*{0.8cm} 10 &   \hspace*{1.5cm} 40,000\\
 Radiologist 2 (H108) &   \hspace*{0.8cm} 15 &  \hspace*{1.5cm} 14,000\\
 Radiologist 3 (H108) &  \hspace*{0.8cm} 11 &  \hspace*{1.5cm} 45,000 \\
 Radiologist 1 (HMUH) &   \hspace*{0.8cm} 21 &  \hspace*{1.5cm} 20,000\\
 Radiologist 2 (HMUH) &   \hspace*{0.8cm} 22 &  \hspace*{1.5cm} 15,000\\
 Radiologist 3 (HMUH) &   \hspace*{0.8cm} 14 &  \hspace*{1.5cm} 15,000\\
\hline
 \textbf{Average} &   \hspace*{0.8cm} \textbf{15.5} &  \hspace*{1.5cm} \textbf{25,000}\\
\hline
\end{tabular}}
\end{table}

\setlength\parindent{0pt} \textbf{\textit{Reader study design}}. The reader study was conducted in two sessions. In the first session, participating readers read the CXR scans independently without the VinDr-CXR assistance. During the second session, the readers re-evaluated all CXR scans with the assistance of the VinDr-CXR. Specifically, the radiologists were provided the VinDr-CXR predictions in the form of bounding boxes, which locate abnormalities (see Figure~\ref{fig:vindr-cxr-interface}; Appendices). They considered the model’s prediction and modified the diagnostics. During this process, the readers were blinded to the relevant clinical information such as the original reports and previous medical histories of the patients or other patient records. To maximize human performance, the readers can perform the task on our browser-based viewer with zoom in or out, panning, and many other support tools. The reader study was set up to ensure that all radiologists can view and interpret the CXR studies in an environment similar to their routine workflow in clinical practice. Changes in radiologists’ agreement were then assessed to investigate the impact of the VinDr-CXR assistance.

\subsection{Statistical analysis}

Diagnostic performance metrics, including area under the receiver operating characteristic curve (AUC), sensitivity, specificity, \textit{F1}-score, false-positive rate (FPR), and false-negative rate (FNR), were used to assess the accuracy of the VinDr-CXR for the classification task. For each indicator, 95\% confidence interval (CI) was estimated with bootstrapping (10,000 replications). To evaluate the VinDr-CXR’s ability to detect and localize lung lesions, we used the FROC (the sensitivities of models under different false positive rates as 0.25, 0.5, 1, 2, and 4). Cohen’s Kappa statistics~\cite{cohen1960coefficient} and percentage agreement rate were used to evaluate the level of agreement between the VinDr-CXR system and participating radiologists, as well as to assess the agreement between pairs of radiologists. To assess inter-rater agreement among a group of radiologists, the Fleiss' Kappa~\cite{fleiss1971measuring} score was used. The Kappa values were interpreted as following guidelines~\cite{landis1977measurement}: ($<$ 0.00): poor; (0.00--0.20): none to slight; (0.21--0.40): fair; (0.41--0.60): moderate; (0.61--0.80): substantial; and (0.81--1.00): almost perfect agreement. All statistical analyses were  performed using Python (version 3.9.2 -- \url{https://www.python.org/}) and scikit-learn (version 0.24.2 -- \url{https://scikit-learn.org/}).

\section{Experimental Results}

This section summarizes our main findings in this study. We first report the performance of the VinDr-CXR on the per-radiograph classification of common lung disease tasks on the internal and external test cohorts. We then provide experimental results for the lesion-level localization task. Finally, we show quantitative assessments on the impact of the VinDr-CXR in clinical practice. 

\subsection{Evaluation of VinDr-CXR performance}

The following subsections detail the quantitative results of the VinDr-CXR system for the classification of diseases and detection of critical lesions on internal and external validation datasets.\\[0.03cm]
\subsubsection*{VinDr-CXR provides accurate per-radiograph classification of common lung disease} 
\begin{figure*}
\centering
\includegraphics[width=18cm,height=11.5cm]{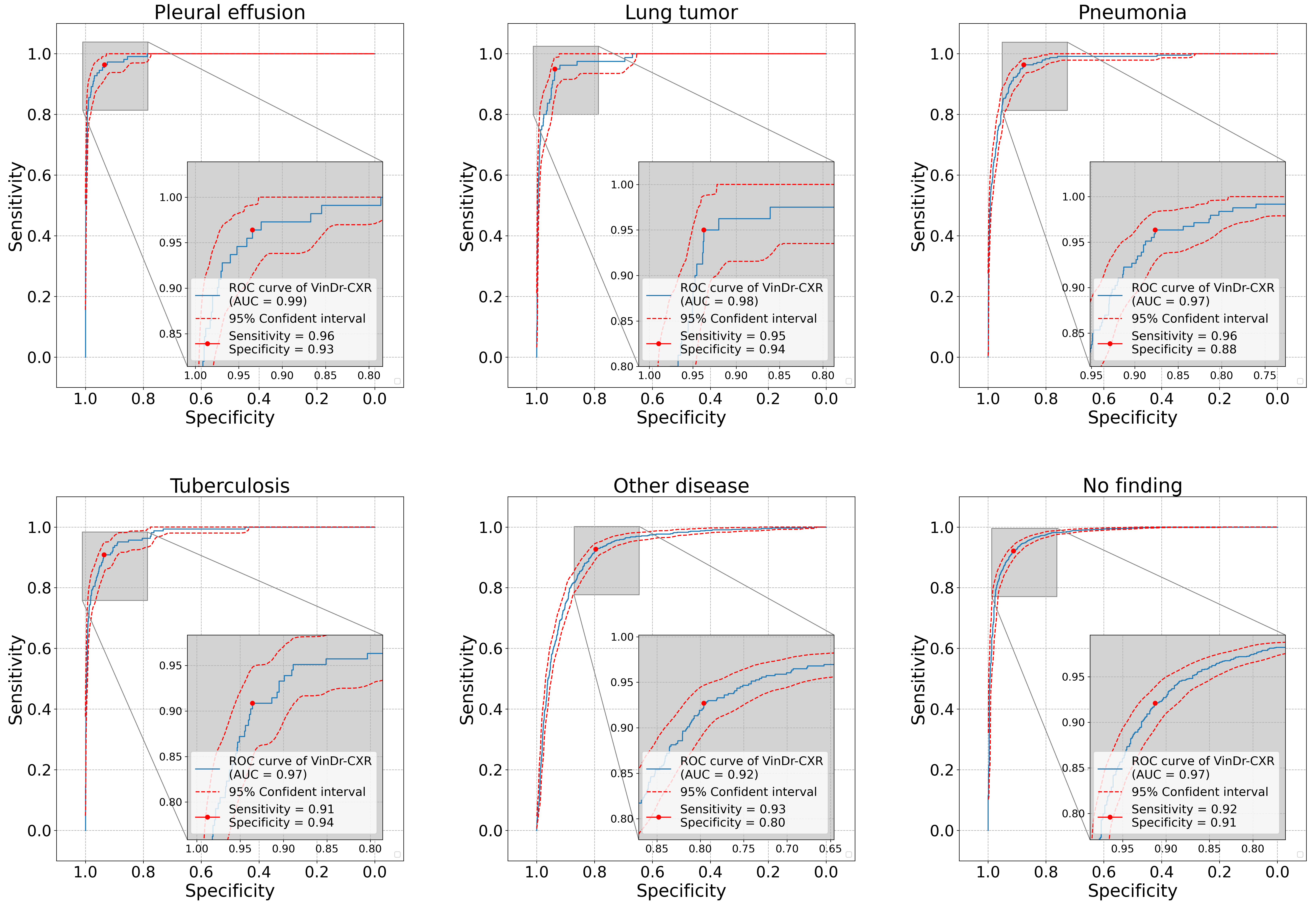}
\caption{Receiver operating curves  (ROC) of the VinDr-CXR system on the internal validation cohort. The solid blue lines show the ROC curves of the system and the dashed red lines show the 95 percentile intervals of the curves based on 10,000 bootstrap samples. We determined the optimal threshold for the VinDr-CXR system  by maximizing Youden's index~\cite{youden1950index} for each disease label. We observe that the DL system achieved consistently high classification performances across all the target diseases (range AUROC: 0.920 -- 0.989).}
\label{fig:roc}
\end{figure*}

The performance of the VinDr-CXR for the classification of common lung diseases was assessed on the internal validation set of 3,000 CXR studies, in which there were 948 patients with abnormal findings or diseases and 2052 patients without any pathologies (Table~\ref{data-characs}). The system achieved a mean AUROC of 0.967 (95\% CI: 0.958, 0.975) over six global disease labels: 0.989 (0.983, 0.994) for Pleural Effusion, 0.978 (0.965, 0.988) for Lung Tumor, 0.969 (0.959, 0.978) for Pneumonia, 0.975 (0.964, 0.983) for Tuberculosis, and 0.920 (0.909, 0.931) for Other Diseases. The sensitivity, specificity, and \textit{F1}-score of the VinDr-CXR were 0.933 (0.898, 0.964), 0.900 (0.887, 0.911), and 0.631 (0.589, 0.672), respectively. The system showed a FPR of 0.101 (0.089, 0.114) and a FNR of 0.067 (0.057, 0.102) over all target diseases. The overall accuracies of the system in differentiating between normal and abnormal CXRs were 0.972 (0.966, 0.978) in AUROC and 0.939 (0.931, 0.947) in \textit{F1}-score. Our experimental results on CXR data from the internal validation set showed high sensitivity and specificity in classifying six disease labels. Detailed performances for individual diseases over all evaluation metrics on the internal validation cohort are reported in Table~\ref{classification_result}. Figure~\ref{fig:roc} shows six ROC curves of the system on the internal validation set for six global diseases.

\begin{table*}
\centering
\textsf{
\scriptsize{
\caption{Per-radiograph classification performance of the VinDr-CXR on the internal validation set (\textit{N = 3,000}). AUC = Area under the receiver operating characteristic curve, FPR = False-positive rate or false alarm rate, FNR = False-negative rate.}
\label{classification_result}
\setlength{\tabcolsep}{3pt}
\begin{tabular}{p{55pt}|p{68pt}|p{68pt}|p{68pt}|p{68pt}|p{68pt}|p{68pt}}
\hline
\hline
 \textbf{Label} &  \textbf{AUC} &  \textbf{Sensitivity} &  \textbf{Specificity} &  \textit{\textbf{F1}}\textbf{-score} &  \textbf{FPR} &  \textbf{FNR}\\
\hline
 Pleural Effusion   &  0.989 (0.983, 0.994) &  0.955 (0.912, 0.991) &  0.934 (0.925, 0.943) &  0.524 (0.463, 0.582) &  0.066 (0.057, 0.075) &  0.045 (0.009, 0.088) \\
 Lung Tumor         &  0.978 (0.965, 0.988) &  0.937 (0.877, 0.986) &  0.937 (0.928, 0.946) &  0.448 (0.381, 0.512) &  0.063 (0.054, 0.072) &  0.063 (0.014, 0.123) \\
 Pneumonia          &  0.969 (0.959, 0.978) &  0.959 (0.933, 0.982) &  0.877 (0.864, 0.889) &  0.576 (0.535, 0.616) &  0.123 (0.111, 0.136) &  0.041 (0.018, 0.067) \\
 Tuberculosis       &  0.975 (0.964, 0.983) &  0.903 (0.854, 0.945) &  0.936 (0.927, 0.945) &  0.602 (0.550, 0.651) &  0.064 (0.055, 0.073) &  0.097 (0.055, 0.146) \\
 Other Diseases     &  0.920 (0.909, 0.931) &  0.925 (0.904, 0.945) &  0.796 (0.780, 0.812) &   0.698 (0.674, 0.722) &  0.204 (0.188, 0.220) &  0.075 (0.055, 0.096) \\
 No Finding         &  0.972 (0.966, 0.978) &  0.920 (0.908, 0.932) &  0.913 (0.895, 0.930) &  0.939 (0.931, 0.947) &  0.087 (0.070, 0.105) &  0.080 (0.068, 0.092) \\
\hline
 \textbf{ Mean}               &   \textbf{0.967 (0.958, 0.975)}          &   \textbf{0.933 (0.898, 0.964)}        &   \textbf{0.900 (0.887, 0.911)}          &   \textbf{0.631 (0.589, 0.672)} &  \textbf{0.101 (0.089, 0.114)}           &   \textbf{0.067 (0.057, 0.102)}  \\
\hline
\hline
\end{tabular}}}
\end{table*}

\subsubsection*{VinDr-CXR shows robust classification performance on external test cohorts}

To investigate the consistency of the VinDr-CXR performance across multiple populations, we performed external validation tests using two independent datasets, including CheXpert~\cite{irvin2019chexpert} and CheXphoto~\cite{phillips2020chexphoto}. The AUROC score, sensitivity, specificity, \textit{F1}-score, FPR, and FNR of the VinDr-CXR with no additional training cost on 200 studies [normal: 27 cases, abnormal: 173 cases] of the CheXpert~\cite{irvin2019chexpert} validation set were 0.892 (0.839, 0.939), 0.866 (0.792, 0.932), 0.614 (0.556, 0.672), 0.453 (0.3650, 0.536), 0.386 (0.328, 0.444), and 0.134 (0.067, 0.208), respectively. On the CheXphoto~\cite{phillips2020chexphoto} validation set [\textit{N = 200}, normal: 27 cases, abnormal: 173 cases], the VinDr-CXR achieved an AUC of 0.888 (0.832, 0.936), a sensitivity of 0.871 (0.800, 0.936), a specificity of 0.614 (0.557, 0.670), a \textit{F1}-score of 0.460 (0.372, 0.542), a FPR of 0.386 ( 0.330, 0.444), and a FNR of 0.129 (0.065, 0.200). The performance of the VinDr-CXR over all disease labels is reported in Table~\ref{tab:external_test}. Although a slight drop has been observed, the performance of the VinDr-CXR on external test sets remained at a high level, showing its robustness across different patient cohorts. These experimental results show evidence that training a DL system with a large-scale, high-quality dataset could reach a high diagnostic accuracy across populations without additional training cost.

\begin{table*}
\centering
\textsf{
\scriptsize{
\caption{Per-radiograph classification performance of the VinDr-CXR on Pleural Effusion, Pneumonia and No Finding from the external validation sets CheXpert~\cite{irvin2019chexpert} (\textit{N = 200})  and CheXphoto~\cite{phillips2020chexphoto} (\textit{N = 200}) datasets.}
\label{tab:external_test}
\setlength{\tabcolsep}{3pt}
\begin{tabular}{p{65pt}|p{65pt}|p{65pt}|p{65pt}|p{65pt}|p{65pt}|p{65pt}}
\hline
\hline
 \textbf{Dataset\&Label} &  \textbf{AUC} &  \textbf{Sensitivity} &  \textbf{Specificity} &  \textit{\textbf{F1}}\textbf{-score} &  \textbf{FPR} &  \textbf{FNR}\\
\hline
 \textbf{CheXpert~\cite{irvin2019chexpert}}  &   &  &  &  &  &  \\
\hline
 Pleural Effusion &   0.895 (0.850, 0.934) &  0.940 (0.877, 0.987) &  0.727 (0.657, 0.795) &  0.715 (0.637, 0.788) &  0.273 (0.205, 0.343) &  0.060 (0.013, 0.123) \\
 Pneumonia        &  0.891 (0.824, 0.949) &  1.000 (1.000, 1.000) &  0.248 (0.193, 0.306) &  0.085 (0.033, 0.143) &  0.752 (0.694, 0.807) &  0.000 (0.000, 0.000) \\
 No Finding       &  0.891 (0.843, 0.933) &  0.659 (0.500, 0.810) &  0.868 (0.818, 0.914) &  0.559 (0.424, 0.678) &  0.132 (0.086, 0.182) &  0.341 (0.190, 0.500) \\
\hline
 Mean             &  0.892 (0.839, 0.939)            &  0.866 (0.792, 0.932)             &  0.614 (0.556, 0.672)             &  0.453 (0.365, 0.536)           &  0.386 (0.328, 0.444)            &  0.134 (0.067, 0.208)            \\
\hline
 \textbf{CheXphoto~\cite{phillips2020chexphoto}}  &  &  &  &  &  &   \\
\hline
 Pleural Effusion &  0.889 (0.843, 0.930) &  0.955 (0.899, 1.000) &  0.739 (0.670, 0.804) &  0.731 (0.654, 0.802) &  0.261 (0.196, 0.331) &  0.045 (0.000, 0.101) \\
 Pneumonia        &  0.887 (0.816, 0.947) &  1.000 (1.000, 1.000) &  0.230 (0.176, 0.288) &  0.083 (0.032, 0.140) &  0.770 (0.712, 0.824) &  0.000 (0.000, 0.000) \\
 No Finding       &  0.887 (0.836, 0.930) &  0.658 (0.500, 0.808) &  0.873 (0.824, 0.919) &  0.566 (0.430, 0.684) &  0.127 (0.081, 0.176) &  0.342 (0.194, 0.500) \\
\hline
 Mean             &   0.888 (0.832, 0.936)            &  0.871 (0.800, 0.936)            &  0.614 (0.557, 0.670)             &   0.460 (0.372, 0.542)         &  0.386 (0.330, 0.444)             &  0.129 (0.065, 0.200)            \\
\hline
\end{tabular}}}
\scriptsize{Abbreviations: AUC = Area under the receiver operating characteristic curve, FPR = False-positive rate or false alarm rate, FNR = False-negative rate or miss detection rate.}
\end{table*}

\subsubsection*{VinDr-CXR provides accurate lesion-level localization}

\begin{figure*}
\centering
\includegraphics[width=13cm,height=15.5cm]{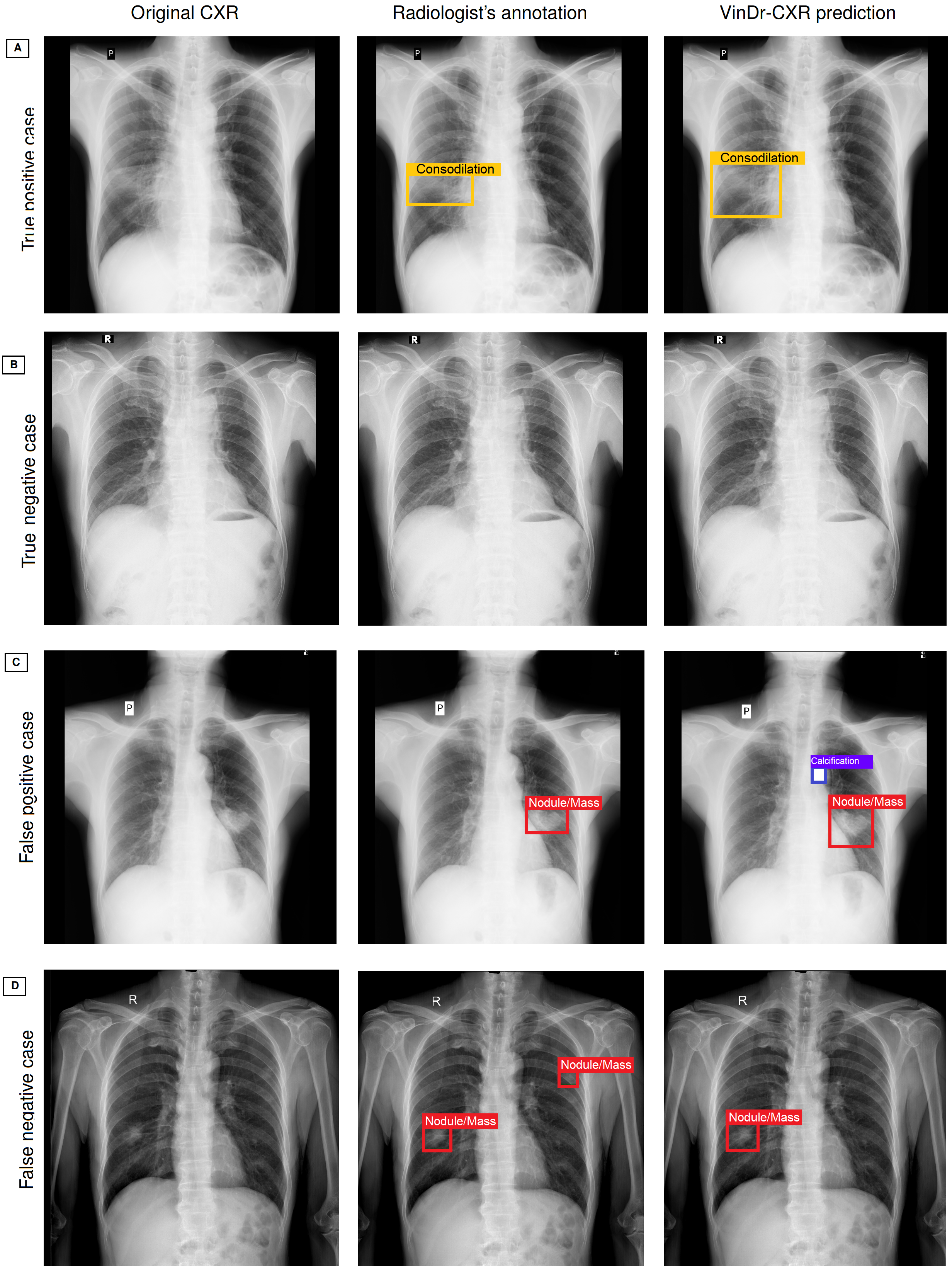}
\caption{Some accurate and erroneous predictions of the VinDr-CXR on several representative CXR images from the internal validation set. (\textbf{A}) The VinDr-CXR correctly identified four lesions, including Pleural Thickening, Calcification, Aortic Enlargement, and Cardiomegaly on the scan. (\textbf{B}) The system correctly identified a normal scan or true negative case. (\textbf{C}) The VinDr-CXR correctly identified a Nodule/Mass. However, it missed another Nodule/Mass on the left lung. (\textbf{D}) The VinDr-CXR correctly identified Nodule/Mass, but it seemed wrong to detect Calcification.}
\label{fig:detection_visualize}
\end{figure*}

\begin{figure*}
\centering
\includegraphics[width=17.5cm,height=13cm]{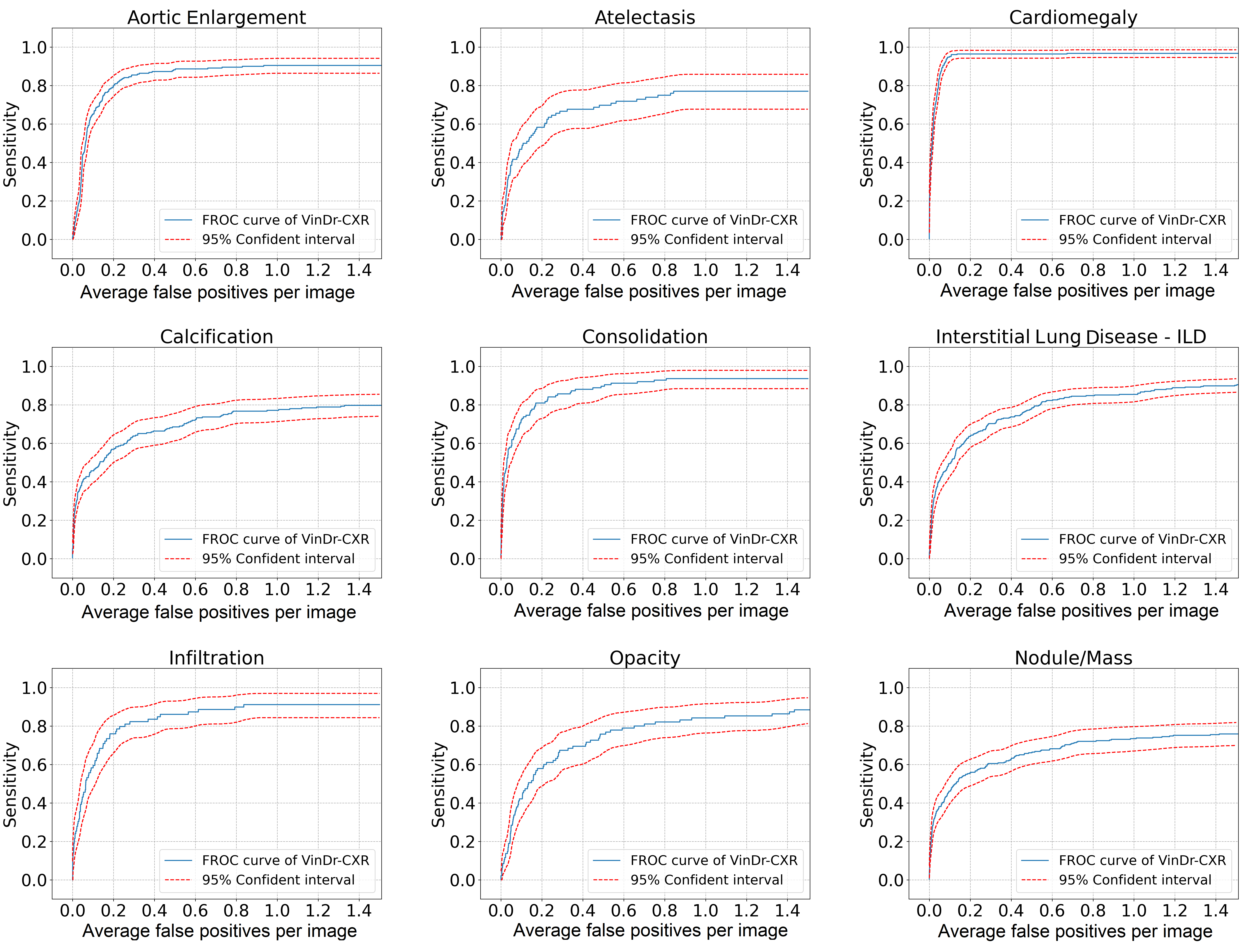}
\caption{Free-response receiver operating characteristic (FROC) of the VinDr-CXR detector for several findings on the internal validation set. Sensitivity is calculated on a per lesion basis. The blue curve shows the results using all validation images, while the red dashed lines show the 95 percentile intervals from 10,000 bootstrap samples.}
\label{fig:froc_curve}
\end{figure*}

For the per-lesion localization task, the VinDr-CXR’s ability to detect and localize abnormal findings was evaluated on 3,000 CXR scans from the internal validation set using FROC analysis~\cite{bunch1978free}. In this experiment, a detection is considered a true positive if the detected bounding box overlaps with the corresponding ground-truth bounding box more than 40\% using the intersection over union (IoU) metric. Otherwise, it is considered to be a false positive. As shown in Table~\ref{tab:detection_performance}, the proposed system achieved a sensitivity of 80.2\% (81.4, 84.9) at 1.0 false-positive marks per image. The FROC (average recall rate at the false positives as 0.25, 0.5, 1.00, 2.00, and 4.00) of the VinDr-CXR system was 78.36\% (76.46, 80.16). FROC curves, which show the sensitivity of the system as a function of the number of false positives marks per image, of some representative findings are shown in Figure~\ref{fig:froc_curve}.

\begin{table*}
        \centering
        \textsf{\scriptsize{
        \caption{Abnormality detection performance of the VinDr-CXR on the internal validation dataset.}
        \label{tab:detection_performance}
        \setlength{\tabcolsep}{2pt}
        \begin{tabular}{
        >{}p{80pt}|
        >{}p{80pt}|
        >{}p{80pt}|
        >{}p{80pt}|
        >{}p{80pt}|
        >{}p{80pt}
        >{}p{80pt}
        }
        \hline
        \hline
        \textbf{Finding} & 
        \textbf{Sensitivity@0.25} & 
        \textbf{Sensitivity@0.5} & 
        \textbf{Sensitivity@1.0} & 
        \textbf{Sensitivity@2.0} & 
        \textbf{Sensitivity@4.0} \\
        \hline
        Cardiomegaly                    & 0.965 (0.943, 0.984) & 0.965 (0.943, 0.984) & 0.968 (0.947, 0.986) & 0.968 (0.947, 0.986) & 0.968 (0.947, 0.986) \\
        Opacity                         & 0.617 (0.517, 0.718) & 0.756 (0.660, 0.848) & 0.842 (0.764, 0.915) & 0.895 (0.827, 0.956) & 0.906 (0.848, 0.959) \\
        Consolidation                   & 0.841 (0.764, 0.912) & 0.898 (0.834, 0.954) & 0.937 (0.884, 0.980) & 0.937 (0.884, 0.980) & 0.937 (0.884, 0.980) \\
        Atelectasis                     & 0.642 (0.538, 0.747) & 0.698 (0.596, 0.796) & 0.772 (0.678, 0.859) & 0.772 (0.678, 0.859) & 0.772 (0.678, 0.859) \\
        Pneumothorax                    & 0.639 (0.467, 0.800) & 0.680 (0.522, 0.828) & 0.680 (0.522, 0.828) & 0.680 (0.522, 0.828) & 0.680 (0.522, 0.828) \\
        Pleural Effusion                & 0.898 (0.840, 0.950) & 0.927 (0.877, 0.970) & 0.934 (0.891, 0.972) & 0.942 (0.902, 0.977) & 0.942 (0.902, 0.977) \\
        Aortic Enlargement              & 0.838 (0.789, 0.885) & 0.882 (0.839, 0.923) & 0.905 (0.864, 0.941) & 0.905 (0.864, 0.942) & 0.909 (0.870, 0.945) \\
        ILD & 0.664 (0.606, 0.723) & 0.782 (0.730, 0.833) & 0.858 (0.816, 0.899) & 0.923 (0.890, 0.954) & 0.937 (0.905, 0.964) \\
        Infiltration                    & 0.801 (0.714, 0.884) & 0.861 (0.787, 0.930) & 0.911 (0.843, 0.969) & 0.911 (0.843, 0.969) & 0.911 (0.843, 0.969) \\
        Nodule/Mass                     & 0.579 (0.510, 0.647) & 0.663 (0.601, 0.727) & 0.735 (0.670, 0.796) & 0.769 (0.710, 0.827) & 0.773 (0.714, 0.830) \\
        Pulmonary Fibrosis              & 0.568 (0.514, 0.623) & 0.627 (0.576, 0.678) & 0.707 (0.657, 0.757) & 0.775 (0.729, 0.819) & 0.796 (0.751, 0.839) \\
        Pleural Thickening              & 0.494 (0.425, 0.564) & 0.608 (0.540, 0.675) & 0.714 (0.651, 0.776) & 0.797 (0.739, 0.852) & 0.850 (0.797, 0.900) \\
        Calcification                   & 0.598 (0.527, 0.670) & 0.685 (0.613, 0.755) & 0.774 (0.713, 0.833) & 0.802 (0.744, 0.858) & 0.802 (0.744, 0.858) \\
        Other Lesions                   & 0.265 (0.197, 0.338) & 0.311 (0.238, 0.388) & 0.372 (0.299, 0.448) & 0.484 (0.408, 0.562) & 0.551 (0.474, 0.627) \\
        No Finding                      & 0.921 (0.909, 0.933) & 0.921 (0.909, 0.933) & 0.921 (0.909, 0.933) & 0.921 (0.909, 0.933) & 0.921 (0.909, 0.933) \\
        \hline
        \textbf{Mean}                            & \textbf{0.689 (0.668, 0.710)} & \textbf{0.751 (0.731, 0.770)} & \textbf{0.802 (0.784, 0.819)} & \textbf{0.832 (0.814, 0.849)} & \textbf{0.844 (0.826, 0.860)} \\
        \hline
\end{tabular}}}
\begin{flushleft} \scriptsize{Model performance was evaluated using the FROC score at different the number of false-positive predictions per image (0.25, 0.5, 1.0, 2.0 and 4.0). Data in parentheses are 95\% confidence intervals.} \end{flushleft}
\end{table*}

\subsubsection*{Anomaly detection visualization}

The per-lesion detection performance of the VinDr-CXR system can be inspected visually through Figure~\ref{fig:detection_visualize}. We investigated the characteristics and errors of the VinDr-CXR detector by visualizing several representative cases containing both correctly detected lesions and lesions that the DL system missed. In this experiment, we used the VinDr-CXR at a sensitivity of 0.802 and 1.0 false-positive marks per image to generate predictions. We found that the system was able to correctly identified almost all critical lesions. Meanwhile, most false-positive detections were small and non-dangerous lung lesions such as calcifications. 

\subsection{Assessment of VinDr-CXR performance in clinical practice}

\subsubsection*{Inter-agreement among radiologists with and without VinDr-CXR assistance}

We conducted a multi-reader study at the HMUH and H108 to investigate the impact of the proposed DL system on radiologists' performance in the regular clinical workflow. At each hospital, three experienced radiologists assessed a total of 200 CXR images without VinDr-CXR assistance at the first read. The images were read with the assistance of VinDr-CXR in the second read. The agreement between the VinDr-CXR and radiologists, as well as inter-observer agreement among radiologists were assessed using the percentage agreement rate and Kappa statistics. The details of the proposed reader study design are provided in the Methods section. In the first read, concordance among three radiologists on data from the H108 showed percentage agreement rates between 89.9\%--90.7\% and Cohen's Kappa values between 0.501 (0.349, 0.654)--0.540 (0.392, 0.688). Agreement among three H108's radiologists was moderate with a Fleiss' Kappa of 0.529 (0.453, 0.605). In the second read, with the support of VinDr-CXR system, agreement among three raters showed a slightly higher percentage agreement rate. Specifically, percentage agreement rates ranged  between 90.2\%--90.7\% and Cohen's Kappa values between 0.519 (0.351, 0.678)--0.556 (0.412, 0.700) that indicated moderate inter-individual agreement. Agreement among three H108's radiologists was moderate with a Fleiss' Kappa of 0.545 (0.465, 0.625), corresponding to an 3.0\% improvement in  Fleiss' Kappa compared to the first read.

\begin{table*}
    \centering
    \caption{Inter-rater agreement among radiologists at the H108 without the VinDr-CXR assistance.}
    \setlength{\tabcolsep}{2pt}
    \begin{tabular}{
    >{}p{70pt} |
    >{}p{25pt} |
    >{}p{70pt} |
    >{}p{25pt} |
    >{}p{70pt} |
    >{}p{25pt} |
    >{}p{70pt} |
    >{}p{70pt}}
    \hline 
     & 
    \multicolumn{2}{l|}{\hspace*{0.5cm} \textbf{Rater 1 \textit{vs}. Rater 2}} & \multicolumn{2}{l|}{\hspace*{0.5cm} \textbf{Rater 1 \textit{vs}. Rater 3}} & \multicolumn{2}{l|}{\hspace*{0.5cm} \textbf{Rater 2 \textit{vs}. Rater 3}} &  \\ \cline{2-7}
    \multirow{-2}{*}{ \textbf{Findings}} & 
    Agree & \hspace*{0.4cm} Cohen's $\kappa$ & 
   Agree & \hspace*{0.4cm} Cohen's $\kappa$ & 
    Agree & \hspace*{0.4cm} Cohen's $\kappa$ & 
    \multirow{-2}{*}{\hspace*{0.4cm} \textbf{Fleiss'} $\kappa$} \\
    \hline
    Lung Tumor         & 0.920 & 0.562 (0.375, 0.748)   & 0.955 & 0.643 (0.426, 0.859)   & 0.905 & 0.490 (0.296, 0.683)   & 0.552 (0.472, 0.632)  \\
    Pneumonia          & 0.895 & 0.698 (0.580, 0.817)   & 0.875 & 0.543 (0.387, 0.700)   & 0.880 & 0.626 (0.496, 0.756)   & 0.626 (0.546, 0.706)  \\
    Tuberculosis       & 0.945 & 0.674 (0.494, 0.853)   & 0.925 & 0.596 (0.413, 0.778)   & 0.950 & 0.763 (0.623, 0.904)   & 0.681 (0.601, 0.761)  \\
    Other Diseases      & 0.830 & 0.651 (0.544, 0.758)   & 0.810 & 0.595 (0.483, 0.707)   & 0.820 & 0.617 (0.507, 0.727)   & 0.619 (0.539, 0.699)  \\
    No Finding         & 0.945 & 0.886 (0.821, 0.952)   & 0.940 & 0.878 (0.812, 0.945)   & 0.945 & 0.889 (0.825, 0.952)   & 0.884 (0.804, 0.964)  \\
    Aortic Enlargement & 0.915 & 0.662 (0.514, 0.811)   & 0.895 & 0.546 (0.377, 0.716)   & 0.930 & 0.669 (0.509, 0.830)   & 0.625 (0.545, 0.705)  \\
    Atelectasis        & 0.925 & -0.009 (-0.027, 0.008) & 0.980 & -0.008 (-0.019, 0.004) & 0.935 & 0.216 (-0.046, 0.478)  & 0.084 (0.004, 0.164)  \\
    Calcification      & 0.905 & 0.654 (0.512, 0.795)   & 0.870 & 0.255 (0.058, 0.452)   & 0.835 & 0.288 (0.122, 0.453)   & 0.419 (0.339, 0.499)  \\
    Cardiomegaly       & 0.915 & 0.574 (0.393, 0.756)   & 0.965 & 0.810 (0.673, 0.946)   & 0.950 & 0.734 (0.578, 0.891)   & 0.703 (0.623, 0.783)  \\
    Consolidation      & 0.940 & 0.308 (0.026, 0.591)   & 0.940 & 0.506 (0.264, 0.749)   & 0.940 & 0.308 (0.026, 0.591)   & 0.388 (0.308, 0.468)  \\
    ILD                & 0.845 & 0.552 (0.412, 0.692)   & 0.820 & 0.348 (0.189, 0.507)   & 0.775 & 0.196 (0.044, 0.349)   & 0.372 (0.292, 0.452)  \\
    Infiltration       & 0.815 & 0.288 (0.139, 0.436)   & 0.860 & 0.239 (0.052, 0.426)   & 0.865 & 0.562 (0.419, 0.706)   & 0.376 (0.296, 0.456)  \\
    Lung Opacity       & 0.830 & 0.167 (-0.014, 0.347)  & 0.870 & 0.054 (-0.067, 0.175)  & 0.890 & -0.019 (-0.042, 0.005) & 0.072 (-0.008, 0.152) \\
    Nodule/Mass        & 0.895 & 0.512 (0.331, 0.694)   & 0.910 & 0.521 (0.328, 0.715)   & 0.915 & 0.591 (0.417, 0.766)   & 0.541 (0.461, 0.621)  \\
    Pleural Effusion   & 0.930 & 0.793 (0.690, 0.896)   & 0.905 & 0.682 (0.549, 0.815)   & 0.935 & 0.800 (0.697, 0.903)   & 0.760 (0.680, 0.840)  \\
    Pleural Thickening & 0.835 & 0.284 (0.116, 0.451)   & 0.890 & 0.255 (0.037, 0.473)   & 0.865 & 0.458 (0.291, 0.624)   & 0.337 (0.257, 0.417)  \\
    Pneumothorax       & 1.000 & 1.000 (1.000, 1.000)   & 1.000 & 1.000 (1.000, 1.000)   & 1.000 & 1.000 (1.000, 1.000)   & 1.000 (1.000, 1.000)  \\
    Pulmonary Fibrosis & 0.870 & 0.505 (0.340, 0.669)   & 0.890 & 0.514 (0.337, 0.691)   & 0.900 & 0.599 (0.441, 0.756)   & 0.539 (0.458, 0.619)  \\
    Other Lesions       & 0.920 & 0.386 (0.145, 0.627)   & 0.935 & 0.546 (0.327, 0.765)   & 0.925 & 0.476 (0.250, 0.702)   & 0.471 (0.391, 0.551)  \\
    \hline
    \textbf{Mean}              & \textbf{0.899} & \textbf{0.534 (0.389, 0.679) }  & \textbf{0.907} & \textbf{0.501 (0.349, 0.654)}   & \textbf{0.903} & \textbf{0.540 (0.392, 0.688)}  & \textbf{0.529 (0.453, 0.605)}  \\
        \hline
\end{tabular}
\end{table*}

\begin{table*}
    \centering
    \caption{Inter-rater agreement among the H108's radiologists with the VinDr-CXR assistance.}
    \setlength{\tabcolsep}{2pt}
    \begin{tabular}{
    >{}p{70pt} |
    >{}p{25pt} |
    >{}p{70pt} |
    >{}p{25pt} |
    >{}p{70pt} |
    >{}p{25pt} |
    >{}p{70pt} |
    >{}p{70pt}}
    \hline 
     & 
    \multicolumn{2}{l|}{\hspace*{0.5cm} \textbf{Rater 1 \textit{vs}. Rater 2}} & \multicolumn{2}{l|}{\hspace*{0.5cm} \textbf{Rater 1 \textit{vs}. Rater 3}} & \multicolumn{2}{l|}{\hspace*{0.5cm} \textbf{Rater 2 \textit{vs}. Rater 3}} &  \\ \cline{2-7}
    \multirow{-2}{*}{ \textbf{Findings}} & 
    Agree & \hspace*{0.4cm} Cohen's $\kappa$ & 
    Agree & \hspace*{0.4cm} Cohen's $\kappa$ & 
    Agree & \hspace*{0.4cm} Cohen's $\kappa$ & 
    \multirow{-2}{*}{\hspace*{0.4cm} \textbf{Fleiss'} $\kappa$ } \\
    \hline
     Lung Tumor         & 0.915 & 0.525 (0.333, 0.717)  & 0.960 & 0.671 (0.458, 0.884)   & 0.895 & 0.436 (0.239, 0.632)    & 0.524 (0.444, 0.604) \\
    Pneumonia          & 0.920 & 0.778 (0.675, 0.881)  & 0.905 & 0.690 (0.561, 0.818)   & 0.885 & 0.655 (0.529, 0.781)    & 0.709 (0.629, 0.789) \\
    Tuberculosis       & 0.950 & 0.710 (0.541, 0.879)  & 0.930 & 0.646 (0.477, 0.814)   & 0.950 & 0.772 (0.637, 0.907)    & 0.711 (0.631, 0.791) \\
    Other Diseases      & 0.815 & 0.621 (0.511, 0.731)  & 0.810 & 0.594 (0.480, 0.707)   & 0.825 & 0.633 (0.526, 0.740)    & 0.614 (0.534, 0.694) \\
    No finding         & 0.950 & 0.897 (0.834, 0.959)  & 0.945 & 0.889 (0.825, 0.952)   & 0.945 & 0.889 (0.825, 0.952)    & 0.891 (0.811, 0.971) \\
    Aortic Enlargement & 0.920 & 0.687 (0.543, 0.830)  & 0.895 & 0.559 (0.393, 0.726)   & 0.935 & 0.708 (0.559, 0.857)    & 0.650 (0.570, 0.730) \\
    Atelectasis        & 0.935 & 0.125 (-0.098, 0.348) & 0.975 & -0.008 (-0.021, 0.005) & 0.940 & 0.312 (0.033, 0.590)    & 0.185 (0.105, 0.265) \\
    Calcification      & 0.930 & 0.764 (0.646, 0.881)  & 0.835 & 0.192 (0.018, 0.365)   & 0.825 & 0.268 (0.105, 0.430)    & 0.438 (0.358, 0.518) \\
    Cardiomegaly       & 0.925 & 0.639 (0.471, 0.807)  & 0.970 & 0.833 (0.703, 0.963)   & 0.955 & 0.775 (0.635, 0.916)    & 0.745 (0.665, 0.825) \\
    Consolidation      & 0.930 & 0.197 (-0.056, 0.451) & 0.935 & 0.484 (0.243, 0.724)   & 0.935 & 0.211 (-0.055, 0.477)   & 0.320 (0.240, 0.400) \\
    ILD                & 0.845 & 0.552 (0.412, 0.692)  & 0.820 & 0.373 (0.214, 0.533)   & 0.795 & 0.296 (0.137, 0.454)    & 0.411 (0.331, 0.491) \\
    Infiltration       & 0.810 & 0.276 (0.127, 0.425)  & 0.835 & 0.229 (0.059, 0.399)   & 0.885 & 0.646 (0.514, 0.777)    & 0.402 (0.322, 0.482) \\
    Lung Opacity       & 0.830 & 0.167 (-0.014, 0.347) & 0.885 & 0.262 (0.063, 0.462)   & 0.865 & -0.055 (-0.086, -0.023) & 0.131 (0.051, 0.211) \\
    Nodule/Mass        & 0.895 & 0.557 (0.389, 0.726)  & 0.905 & 0.573 (0.401, 0.745)   & 0.920 & 0.647 (0.487, 0.806)    & 0.592 (0.512, 0.672) \\
    Pleural Effusion   & 0.940 & 0.825 (0.730, 0.920)  & 0.925 & 0.759 (0.643, 0.875)   & 0.945 & 0.834 (0.739, 0.928)    & 0.807 (0.727, 0.887) \\
    Pleural Thickening & 0.840 & 0.318 (0.151, 0.485)  & 0.885 & 0.197 (-0.015, 0.410)  & 0.855 & 0.420 (0.253, 0.587)    & 0.321 (0.241, 0.401) \\
    Pneumothorax       & 0.995 & 0.798 (0.410, 1.000)  & 0.995 & 0.798 (0.410, 1.000)   & 1.000 & 1.000 (1.000, 1.000)    & 0.855 (0.775, 0.935) \\
    Pulmonary Fibrosis & 0.865 & 0.518 (0.360, 0.675)  & 0.875 & 0.530 (0.368, 0.691)   & 0.900 & 0.661 (0.525, 0.798)    & 0.571 (0.491, 0.651) \\
    Other Lesions       & 0.920 & 0.386 (0.145, 0.627)  & 0.940 & 0.593 (0.385, 0.801)   & 0.920 & 0.457 (0.233, 0.681)    & 0.482 (0.402, 0.562) \\ 
    \hline
    \textbf{Mean}             & \textbf{0.902} & \textbf{0.544 (0.374, 0.704)}  & \textbf{0.907} & \textbf{0.519 (0.351, 0.678) }  & \textbf{0.904} & \textbf{0.556 (0.412, 0.700)}    & \textbf{0.545 (.465, .625)} \\
        \hline
\end{tabular}
\end{table*}

We observed the same results on the data from the HMUH. Details of clinical evaluation results at the HMUH are provided in Table~\ref{tab:HMUH_agreement_ai} and Table~\ref{tab:HMUH_agreement_with_ai} (Appendices). In the first read, the percentage agreement rates between each pair of radiologists ranged from 90.2\%--90.8\%; Cohen’s Kappa values were between 0.367 (0.192, 0.543)--0.483 (0.304, 0.662), and a Fleiss' Kappa of 0.404 (0.329–0.480). In the second read, percentage agreement rates were between 90.2\%--90.8\%, and Cohen's Kappa values were between 0.367 (0.192, 0.543)--0.483 (0.304, 0.662). The agreement between three HMUH's radiologists was a Fleiss' Kappa of 0.418 (0.342, 0.494), corresponding to an improvement of 3.4\% compared to the first read. 

\subsubsection*{Diagnostic agreement between VinDr-CXR and radiologists with and without assistance}

In this experiment, the rate of agreement between the VinDr-CXR and radiologists in detecting abnormal lung findings from CXRs were assessed. Table~\ref{tab:AI_rad_H108_agreement_no_support} and Table~\ref{tab:AI_rad_H108_agreement_support} show the agreement rate between the VinDr-CXR and H108's radiologists without and with assistance, respectively. In the first read, the percentage agreement rates were 88.8\%--90.2\%, and the Cohen's Kappa values ranged from 0.462 (0.283, 0.640)--0.506 (0.327, 0.685). With the assistance of VinDr-CXR, the rates of agreement ranged from 90.5\%--91.1\% in percentage agreement and from 0.524 (0.348, 0.699)--0.546 (0.370, 0.717) in Cohen's Kappa values. At the HMUH, the percentage agreement rates were 88.8\%--90.2\%, and the Cohen's Kappa values ranged from 0.462 (0.283, 0.640)--0.506 (0.327, 0.685) without the VinDr-CXR assistance. Meanwhile, the rates of agreement ranged from 90.5\%--91.1\% in percentage agreement and from 0.524 (0.348, 0.699)--0.546 (0.370, 0.717) in Kappa values with the VinDr-CXR assistance (Detailed in Table~\ref{tab:AI_rad_HMU_agreement_no_support} and Table~\ref{tab:AI_rad_HMU_agreement_support} in the Appendices). After consulting the VinDr-CXR output, significant improvements in Kappa scores have been observed across two hospitals. These results indicated that the VinDr-CXR assistance resulted in a significant increase in the agreement between the DL system and radiologists.

\begin{table*}
    \centering
    \textsf{\scriptsize{
    \caption{Agreement between the VinDr-CXR and H108's radiologists without assistance.}
    \label{tab:AI_rad_H108_agreement_no_support}
    \setlength{\tabcolsep}{2pt}
    \begin{tabular}{
    >{}p{70pt}|
    >{}p{25pt}|
    >{}p{70pt}|
    >{}p{25pt}|
    >{}p{70pt}|
    >{}p{25pt}|
    >{}p{70pt}
    }
    \hline
     & 
    \multicolumn{2}{l|}{\hspace*{1cm} \textbf{Rater 1 \textit{vs}. AI}} & 
    \multicolumn{2}{l|}{\hspace*{1cm} \textbf{Rater 2 \textit{vs}. AI}} & 
    \multicolumn{2}{l }{\hspace*{1cm} \textbf{Rater 3 \textit{vs}. AI}} \\ \cline{2-7}
    \multirow{-2}{*}{\textbf{Findings}} & 
    Agree &\hspace*{0.6cm} Cohen's $\kappa$ & 
    Agree &\hspace*{0.6cm} Cohen's $\kappa$ & 
    Agree &\hspace*{0.6cm} Cohen's $\kappa$ \\
    \hline
    Lung Tumor         & 0.945 & 0.493 (0.235, 0.751)  & 0.885 & 0.329 (0.129, 0.530)  & 0.940 & 0.469 (0.215, 0.723)   \\
    Pneumonia          & 0.900 & 0.688 (0.561, 0.815)  & 0.905 & 0.735 (0.623, 0.846)  & 0.875 & 0.566 (0.418, 0.715)   \\
    Tuberculosis       & 0.965 & 0.769 (0.604, 0.934)  & 0.950 & 0.723 (0.560, 0.885)  & 0.940 & 0.695 (0.534, 0.856)   \\
    Other Diseases      & 0.725 & 0.399 (0.277, 0.521)  & 0.705 & 0.355 (0.231, 0.479)  & 0.795 & 0.487 (0.353, 0.620)   \\
    No Finding         & 0.905 & 0.810 (0.730, 0.890)  & 0.900 & 0.800 (0.718, 0.882)  & 0.935 & 0.870 (0.802, 0.938)   \\
    Aortic Enlargement & 0.880 & 0.359 (0.174, 0.544)  & 0.895 & 0.361 (0.159, 0.563)  & 0.915 & 0.378 (0.150, 0.605)   \\
    Atelectasis        & 0.910 & 0.091 (-0.076, 0.259) & 0.895 & 0.308 (0.087, 0.529)  & 0.900 & 0.067 (-0.093, 0.226)  \\
    Calcification      & 0.875 & 0.326 (0.127, 0.525)  & 0.860 & 0.418 (0.252, 0.585)  & 0.875 & 0.008 (-0.136, 0.152)  \\
    Cardiomegaly       & 0.920 & 0.607 (0.432, 0.782)  & 0.905 & 0.542 (0.360, 0.724)  & 0.935 & 0.662 (0.491, 0.832)   \\
    Consolidation      & 0.920 & 0.458 (0.235, 0.681)  & 0.910 & 0.219 (-0.006, 0.444) & 0.930 & 0.526 (0.309, 0.743)   \\
    ILD                & 0.830 & 0.488 (0.340, 0.636)  & 0.845 & 0.537 (0.394, 0.680)  & 0.840 & 0.385 (0.219, 0.550)   \\
    Infiltration       & 0.830 & 0.226 (0.062, 0.391)  & 0.905 & 0.712 (0.592, 0.833)  & 0.870 & 0.530 (0.372, 0.687)   \\
    Lung Opacity       & 0.860 & 0.338 (0.147, 0.528)  & 0.860 & 0.255 (0.058, 0.453)  & 0.880 & -0.019 (-0.043, 0.005) \\
    Nodule/Mass        & 0.860 & 0.383 (0.200, 0.566)  & 0.875 & 0.489 (0.316, 0.661)  & 0.870 & 0.409 (0.225, 0.593)   \\
    Pleural Effusion   & 0.975 & 0.921 (0.853, 0.989)  & 0.945 & 0.839 (0.747, 0.931)  & 0.920 & 0.735 (0.613, 0.857)   \\
    Pleural Thickening & 0.905 & 0.337 (0.109, 0.565)  & 0.850 & 0.390 (0.220, 0.559)  & 0.915 & 0.494 (0.286, 0.701)   \\
    Pneumothorax       & 0.985 & 0.565 (0.125, 1.000)  & 0.985 & 0.565 (0.125, 1.000)  & 0.985 & 0.565 (0.125, 1.000)   \\
    Pulmonary Fibrosis & 0.900 & 0.558 (0.386, 0.730)  & 0.850 & 0.398 (0.224, 0.572)  & 0.870 & 0.384 (0.194, 0.575)   \\
    Other Lesions       & 0.945 & 0.563 (0.332, 0.794)  & 0.955 & 0.643 (0.426, 0.859)  & 0.940 & 0.568 (0.350, 0.787)   \\
    \hline
    \textbf{Mean}               & \textbf{0.897} & \textbf{0.494 (0.308, 0.679)}  & \textbf{0.888} & \textbf{0.506 (0.327, 0.685)}  & \textbf{0.902} & \textbf{0.462 (0.283, 0.640)}  \\
    \hline
    \end{tabular}}}
\end{table*}

\begin{table*}
    \centering
    \caption{Agreement between the VinDr-CXR and H108's radiologists with assistance.}
    \label{tab:AI_rad_H108_agreement_support}
    \setlength{\tabcolsep}{2pt}
    \begin{tabular}{
    >{}p{70pt}|
    >{}p{25pt}|
    >{}p{70pt}|
    >{}p{25pt}|
    >{}p{70pt}|
    >{}p{25pt}|
    >{}p{70pt}
    }
    \hline
     & 
    \multicolumn{2}{l|}{\hspace*{1cm} \textbf{Rater 1 \textit{vs}. AI}} & 
    \multicolumn{2}{l|}{\hspace*{1cm} \textbf{Rater 2 \textit{vs}. AI}} & 
    \multicolumn{2}{l }{\hspace*{1cm} \textbf{Rater 3 \textit{vs}. AI}} \\ \cline{2-7}
    \multirow{-2}{*}{ \textbf{Findings}} & 
    Agree & \hspace*{0.6cm} Cohen's $\kappa$ & 
    Agree & \hspace*{0.6cm} Cohen's $\kappa$ & 
    Agree & \hspace*{0.6cm} Cohen's $\kappa$  \\
    \hline
    Lung Tumor         & 0.950 & 0.519 (0.259, 0.780)  & 0.885 & 0.329 (0.129, 0.530)  & 0.950 & 0.558 (0.313, 0.802)  \\
    Pneumonia          & 0.925 & 0.776 (0.668, 0.884)  & 0.905 & 0.735 (0.623, 0.846)  & 0.910 & 0.703 (0.575, 0.830)  \\
    Tuberculosis       & 0.970 & 0.807 (0.657, 0.957)  & 0.950 & 0.723 (0.560, 0.885)  & 0.940 & 0.707 (0.553, 0.862)  \\
    Other Diseases      & 0.735 & 0.414 (0.291, 0.537)  & 0.700 & 0.356 (0.234, 0.477)  & 0.815 & 0.543 (0.415, 0.670)  \\
    No finding         & 0.910 & 0.820 (0.742, 0.898)  & 0.900 & 0.800 (0.718, 0.882)  & 0.935 & 0.870 (0.802, 0.938)  \\
    Aortic Enlargement & 0.880 & 0.359 (0.174, 0.544)  & 0.900 & 0.408 (0.209, 0.606)  & 0.915 & 0.417 (0.199, 0.635)  \\
    Atelectasis        & 0.910 & 0.091 (-0.076, 0.259) & 0.905 & 0.374 (0.151, 0.597)  & 0.905 & 0.146 (-0.057, 0.349) \\
    Calcification      & 0.905 & 0.550 (0.376, 0.723)  & 0.865 & 0.449 (0.286, 0.612)  & 0.870 & 0.002 (-0.138, 0.142) \\
    Cardiomegaly       & 0.925 & 0.625 (0.451, 0.799)  & 0.920 & 0.634 (0.470, 0.799)  & 0.935 & 0.662 (0.491, 0.832)  \\
    Consolidation      & 0.925 & 0.506 (0.289, 0.722)  & 0.905 & 0.146 (-0.057, 0.349) & 0.940 & 0.594 (0.386, 0.801)  \\
    ILD                & 0.840 & 0.518 (0.372, 0.664)  & 0.845 & 0.537 (0.394, 0.680)  & 0.850 & 0.449 (0.286, 0.611)  \\
    Infiltration       & 0.845 & 0.304 (0.134, 0.475)  & 0.905 & 0.712 (0.592, 0.833)  & 0.890 & 0.627 (0.485, 0.769)  \\
    Lung Opacity       & 0.860 & 0.338 (0.147, 0.528)  & 0.860 & 0.255 (0.058, 0.453)  & 0.905 & 0.308 (0.087, 0.529)  \\
    Nodule/Mass        & 0.875 & 0.489 (0.316, 0.661)  & 0.880 & 0.516 (0.347, 0.685)  & 0.880 & 0.487 (0.311, 0.663)  \\
    Pleural Effusion   & 0.985 & 0.954 (0.901, 1.000)  & 0.945 & 0.839 (0.747, 0.931)  & 0.930 & 0.773 (0.659, 0.886)  \\
    Pleural Thickening & 0.915 & 0.407 (0.178, 0.636)  & 0.855 & 0.420 (0.253, 0.587)  & 0.910 & 0.451 (0.237, 0.664)  \\
    Pneumothorax       & 0.990 & 0.745 (0.405, 1.000)  & 0.985 & 0.565 (0.125, 1.000)  & 0.985 & 0.565 (0.125, 1.000)  \\
    Pulmonary Fibrosis & 0.905 & 0.587 (0.420, 0.754)  & 0.870 & 0.508 (0.347, 0.670)  & 0.900 & 0.599 (0.441, 0.756)  \\
    Other Lesions       & 0.945 & 0.563 (0.332, 0.794)  & 0.955 & 0.643 (0.426, 0.859)  & 0.935 & 0.546 (0.328, 0.764)  \\
    \hline
    \textbf{Mean}              & \textbf{0.905} & \textbf{0.546 (0.370, 0.717)} & \textbf{0.891} & \textbf{0.524 (0.348, 0.699)}  & \textbf{0.911} & \textbf{0.527 (0.342, 0.711)} \\
    \hline
    \end{tabular}
\end{table*}

\subsubsection*{Impact of VinDr-CXR on radiologist diagnostic agreement} 
\begin{figure*}
\centering
\includegraphics[width=18cm,height=6cm]{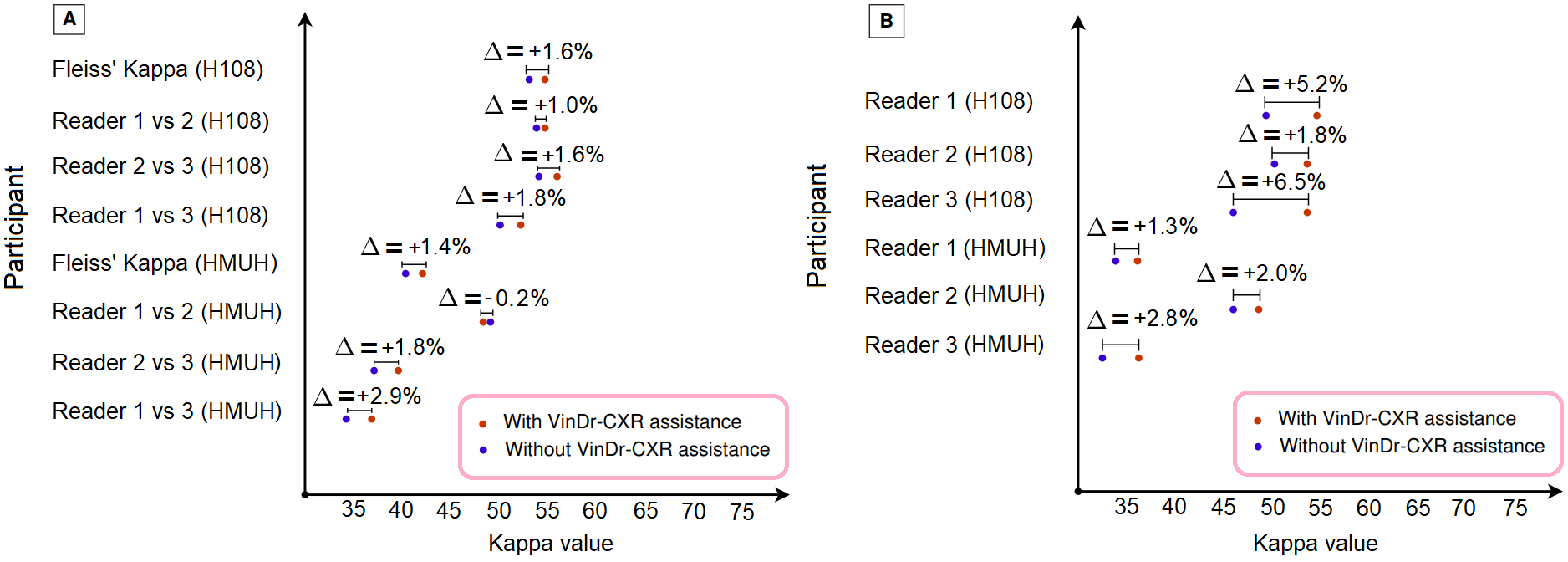}
\caption{\textbf{Impact of the VinDr-CXR on radiologists' performance.} (\textbf{A}) Change in inter-radiologist agreement before and after consulting the VinDr-CXR predictions. The VinDr-CXR assistance significantly improved agreement between radiologists with an increase of 1.5\% in mean Fleiss’ Kappa.  (\textbf{B}) Change in individual radiologists’ agreement with the system before and after consulting the VinDr-CXR predictions.  The VinDr-CXR assistance resulted in a significant increase in the agreement between the AI system and radiologists with an improvement of 3.3\% in mean Cohen’s Kappa. In particular, all differences were statistically significant.}
\label{fig:reader_study}
\end{figure*}

We show evidence that the VinDr-CXR helped improve the degree of agreement among radiologists for the task of detection of lung lesions (see Figure~\ref{fig:reader_study}\textbf{A}). Specifically, Fleiss' Kappa values improved by 1.4\% and 1.6\% for HMUH and H108 readers, respectively. The inter-rater reliability has improved by 1.0\%--2.9\%, except for Reader 1 and Reader 2 from the HMUH ($\Delta$ = -0.2\%). Additionally, we found that the rate of VinDr-CXR agreement with the participating radiologists was slightly higher than the rate of agreement among radiologists. With the support of VinDr-CXR, the agreement between the proposed DL system and radiologists significantly improved. As shown in Figure~\ref{fig:reader_study}\textbf{B}, increments of agreement degree ranged 1.8\%--6.5\% after consulting the VinDr-CXR predictions. Figure~\ref{fig:reader_study_vis} in the Appendices shows several representative images from our reader study, which indicates the change in diagnostic decision after viewing the VinDr-CXR recommendation. 

\subsection*{Visual interpretability of VinDr-CXR}

\begin{figure*}
\centering
\includegraphics[width=17cm,height=4.5cm]{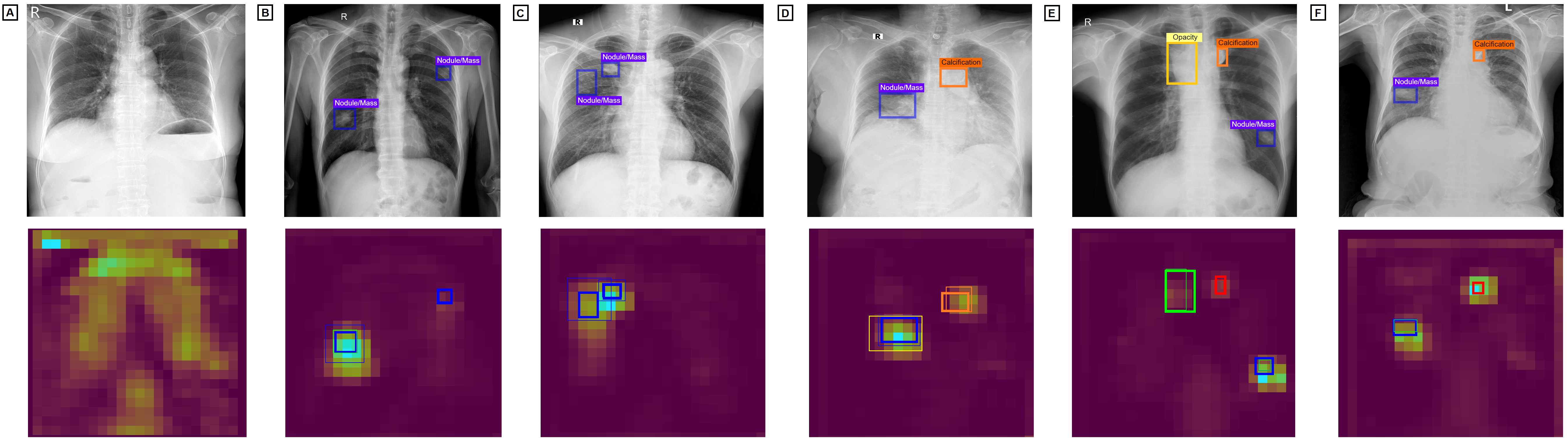}
\caption{Visual explainability of the VinDr-CXR system. Visualization of CXRs and saliency maps overlaid on the original CXR images for several patients chosen from the internal validation set. The CXRs and affected areas were provided by our expert radiologist (top row). The corresponding saliency maps for CXRs were found by the VinDr-CXR classification model (bottom row). The blue, red, and green boxes cover the regions of Nodule/Mass, Calcification, and Opacity, respectively. Boxes with thick and thin lines denote ground-truth boxes and detected boxes, respectively. This figure indicates that the saliency maps highlighted parts of the CXRs that contain abnormal patterns (local labels), which are clinically correlated with disease conclusions (global labels). Best viewed in color.}
\label{fig:vis-xai}
\end{figure*}

Explainability is an absolute necessity for the broad deployment of AI models in clinical practice. In a recent publication, the United States Food and Drug Administration (FDA) mentioned that explainability is required~\cite{FDA} for any AI-based computer-aided diagnosis (CAD) system. In summary, an interpretable AI/DL system can show the links between the features extracted by the system and its predictions~\cite{xie2020chexplain}. Particularly, those links can be understood by a human expert. Explainable DL systems help human experts understand the underlying reasoning of DL systems and identify individual cases in which an AI model potentially gives incorrect predictions. In this study, the proposed VinDr-CXR is not only able to provide disease conclusions (global labels), but a helpful explanation involves abnormal findings (local labels) with their corresponding exact locations. Beyond the classification output, the VinDr-CXR can provide localization information that locates abnormalities accurately on CXR scans. The bounding boxes provided by the DL system may be an essential consideration supporting classification outputs. To illustrate the interpretability of the system, we utilized the trained classification network to compute and visualize the saliency maps for several examples from the internal validation set. To this end, we extracted feature maps produced by the last convolutional layer of the VinDr-CXR classifier model. We then used principal component analysis (PCA)~\cite{karamizadeh2013overview} to reduce the channels of the feature map into a single channel. Then, this single-channel map is converted to a saliency map for visualization. Figure~\ref{fig:vis-xai} shows the original CXRs and lesion bounding boxes annotated by our expert radiologists for Lung Tumor and Tuberculosis (TB) patients. The corresponding saliency maps obtained from the VinDr-CXR system are also provided. We observed the following insights that help better understand the decision-making process of the VinDr-CXR system. First, the visualization of normal cases was irregular with a symmetric high colormap, and there was no increased signal over all parts of the lung (Figure~\ref{fig:vis-xai}\textbf{A}). Second, across all abnormal scans, the saliency maps highlighted parts of the CXRs that contain abnormal patterns such as Nodule/Mass, Calcification, and Opacity, which are clinically correlated with disease conclusions, including Lung Tumor and TB. In other words, the attention regions in the visualization maps were consistent with the annotated abnormal findings provided by our radiologists, as well as  predictions by the VinDr-CXR detector model (Figure~\ref{fig:vis-xai}\textbf{B--F}).

\section{Discussion}

The VinDr-CXR may find helpful in several clinical scenarios. (1) The system was able to discriminate correctly between normal and abnormal CXRs. It, therefore, could be used as a tool to automate screening of common lung diseases for primary diagnosis (e.g., Tuberculosis and Malignancies) at scale. (2) Non-specialist clinicians could also potentially use the VinDr-CXR at healthcare centers to provide teleradiology primary CXR reading or to support rapidly triage cases. (3) The VinDr can be used as a second-reader or assistance tool for clinicians. In this setting, the system can provide diagnosis assistance at different levels including image-level diagnosis and lesion-level diagnosis. This makes the proposed DL algorithms more transparent and more explainable, allowing clinicians to understand and explore the results to improve diagnosis outcomes.\\[0.3cm]
On the technical aspect, this study strongly supports the following statements. First, a DL network can learn effectively and accurately predict common thoracic diseases and findings if trained on large-scale, multi-institutional, and expert-annotated imaging datasets. Second, although the current study was conducted in Vietnam, the proposed VinDr-CXR showed its robustness on external datasets acquired from hospitals in the United States. This result showed evidence that the model generalization of a DL system could consistent across geographical settings. Several studies~\cite{zech2018variable,kelly2019key} showed that DL models for the CXR interpretation failed to generalize to image sources from new institutions and hospitals. In contrast, our finding is similar to Cohen \textit{et al.}~\cite{cohen2020limits} who presented evidence that generalization over difference distributions is not due to a shift in the images but instead a shift in the labels. Third, our external validation experiment showed evidence that a DL system trained on digital CXR could generalize well on CXR captured by smartphone cameras without additional training cost. This opens the opportunity to integrate the DL system like VinDr-CXR into the vast spectrum of clinical workflows across the world, including developing regions that is still using films.\\[0.3cm]
In terms of novelty, the VinDr-CXR system shows several contributions. First, while most previous studies~\cite{rajpurkar2017chexnet,rajpurkar2018deep,irvin2019chexpert,majkowska2020chest,rajpurkar2020chexpedition,tang2020automated} used machine-generated annotations that contain many CXR images with uncertainty labels to train DL systems, our model was trained on radiologist-generated annotations for both development and validation data sets. Next, the proposed DL system evaluated both the detection and classification tasks, while most previous studies have only evaluated image-level classification performance without specifying the location of abnormal findings. Last, this study conducted a large-scale, clinical evaluation to investigate the actual impact of a DL system on the variability in radiologist performance in the interpretation of CXRs.  We showed that the system significantly improved agreement among radiologists. To the best of our knowledge, we are the first to show that a DL system trained on a large-scale, annotated dataset can offer clinical value by helping to improve the rate of agreement among physicians. Furthermore, we also observed that VinDr-CXR assistance resulted in a significant increase in the agreement between the DL system and radiologists. Note that most literature refers to comparisons between human performing and AI models on the CXR interpretation, usually in diagnostic accuracy on the same clinical validation dataset. We suggest that these comparisons do not offer valuable insights into these systems' impact on clinical practice.\\[0.3cm]
This study is not without limitations. First, the development and evaluation datasets only contain frontal CXR scans. Meanwhile, several clinical findings require lateral views. The next version of the dataset may consider adding the lateral views to train the VinDr-CXR system. Second, in clinical practice, physicians diagnose diseases based on both the patient’s clinical history and visual information from CXRs. The VinDr-CXR, however, used only image information for providing diagnosis results without taking clinical and laboratory information into account. Third, the current study's most significant limitation is that we did not directly measure the actual impact of the VinDr-CXR on the sensitivity and specificity of participating radiologists due to the lack of gold reference ground truth. We showed in this study that the DL system was able to reduce clinical disagreements among radiologists. However, there is no clear evidence that the DLS helps improve the sensitivity or specificity of the radiologist in CXR interpretation.

\section{Conclusion}

We reported in this paper the development and validation of an AI-based system called VinDr-CXR for identifying 14 abnormal findings and classifying six common lung diseases from CXRs. To achieve this goal, we collected and annotated a large-scale CXR dataset of 53,485 studies across two major hospitals. Using radiologist-generated annotations as the reference standard to train the VinDr-CXR, we showed that it could achieve a performance level on par with a group of experienced radiologists in classifying common thoracic diseases. The system also achieved high diagnostic accuracy for most abnormal findings. Extensive validation experiments confirmed that the VinDr-CXR system generalizes well to datasets acquired  from different patient cohorts. Most importantly, our clinical evaluations determined the value of the VinDr-CXR in clinical practice when it improved the agreement among physicians. The proposed system could be directly applied in different clinical settings, e.g., supporting physicians in triaging cases or using as a second reader. Although many DL–based models for predicting lung diseases have improved diagnostic accuracy, in some cases surpassing radiologists’ performance, there is little evidence showing that deployment of these models has improved patient outcomes. Therefore, further research is needed to validate the model prospectively and determine its utility in clinical settings. For example, the diagnostic and clinical effects of the VinDr-CXR needs to be assessed in large-scale test cohorts to determine the change in sensitivity and specificity of radiologists for the CXR interpretation in routine clinical practice.

\section*{Appendices}

\subsection*{Development of deep learning algorithms}

\begin{figure*}
\centering
\includegraphics[width=13.5cm,height=4cm]{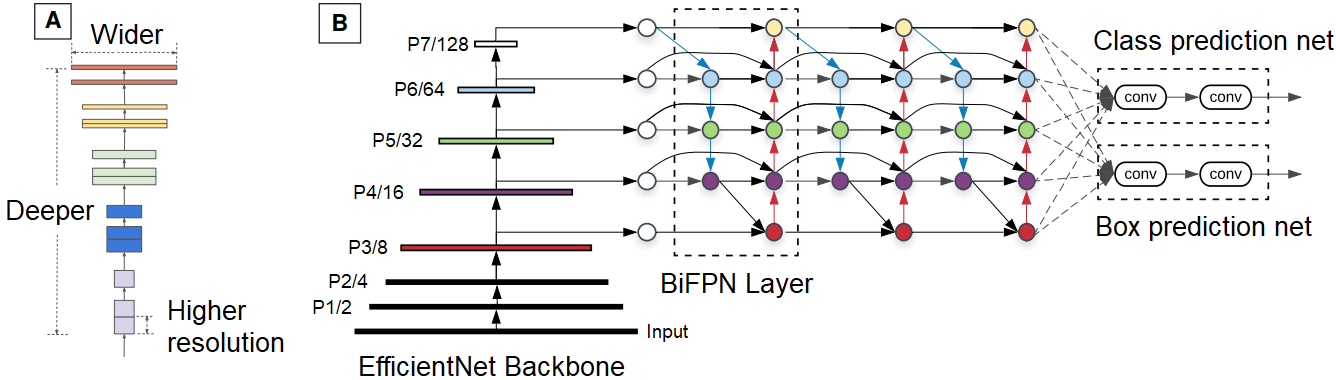}
\caption{(\textbf{A}) EfficientNet architecture: scaling method that uniformly scales all three dimensions with a fixed ratio. (\textbf{B}) EfficientDet architecture employs EfficientNet as the backbone network, BiFPN as the feature network, and shared class/box prediction network. These figures were reproduced from the original papers.}
\label{fig:networks}
\end{figure*}

\textbf{\textit{Network architectures}}. Deep learning (DL)~\cite{lecun2015deep}, a subfield of machine learning, is a computational model that composes multiple processing layers and uses data-driven rules to learn representations of data with multiple levels of abstraction. DL networks showed their breakthrough successes in a wide variety of diagnostic tasks in medical imaging analysis~\cite{litjens2017survey,rajpurkar2017chexnet,esteva2017dermatologist,wu2019deep}. In this study, we applied two well-known DL networks for classifying common thoracic diseases and detecting abnormal findings in CXR images. We deployed EfficientNet-B6~\cite{tan2019efficientnet} for the task of disease classification and EfficientDet-D6~\cite{tan2020efficientdet} for the task of lesion detection. These network architectures were well-known as the most commonly used and most successful DL networks for image classification and object detection. By balancing network depth, width, and resolution, the EfficientNet~\cite{tan2019efficientnet} can lead to much better accuracy and efficiency than other state-of-the-art CNN models. Meanwhile, the EfficientDet-D6~\cite{tan2020efficientdet} used a weighted bi-directional feature pyramid network (BiFPN), allowing easy and fast multiscale feature fusion and combined with the feature learning capacity of EfficientNet-B6~\cite{tan2019efficientnet}. This network architecture design, which requires less computational resources for training, achieved much better efficiency than prior state-of-the-art detectors. Therefore, we found that EfficientNet-B6~\cite{tan2019efficientnet} and EfficientDet-D6~\cite{tan2020efficientdet} well-suited for medical applications, including the CXR analysis. Figure~\ref{fig:networks} illustrates the key ideas behind these two network architectures. 
\\
\\
\textbf{\textit{Implementation details}}. We followed the original implementation of EfficientNet-B6~\cite{tan2019efficientnet} and EfficientDet-D6~\cite{tan2020efficientdet} with some minor modifications. Specifically, the last fully-connected layer of EfficientNet-B6~\cite{tan2019efficientnet} has changed to output a vector of six dimensions corresponding to six classes. Mean binary cross-entropy loss was used to train these six classes simultaneously instead of a single multi-class cross-entropy loss in the original works. For the detector, we only changed the classification head to return scores for 14 lesion types while the loss functions and other parts of the architecture were preserved.\\[0.3cm]
\textbf{\textit{Data augmentation}.} Several pre-processing steps were performed on CXRs before passing to the networks. Images in the DICOM format were converted to 8-bit PNG, then padded and resized to $1024\times1024$ pixels. Subsequently, 1-channel images were transformed to 3-channel ones by repeating the channel three times. In the last step, images with the pixels intensity in range $[0, 255]$ were normalized by subtracting $(123.675, 116.28, 103.53)$ then dividing by $(58.395, 57.12, 57.375)$ in a channel-wise manner.\\[0.3cm]
\textbf{\textit{Training procedures}.} Regarding the training procedure, the classifier’s weights were initialized with weights trained on the ImageNet dataset~\cite{krizhevsky2012imagenet}, a large-scale natural image dataset for the classification task. Pre-processed training images were randomly transformed using resize-cropping, shift-scale-rotating, horizontal flipping, brightness-contrast adjustments, then grouped into batches of 16. The classifier was trained for multi-label binary classification tasks (presence or absence of each disease) by optimizing the mean binary cross-entropy loss of all diseases. A variant of the stochastic gradient descent algorithm, Adam optimizer~\cite{kingma2014adam}, was used with a base learning rate of $2\times10e-4$. The learning rate was then linearly increased in the first epoch then gradually annealed to 0 at the end of the 50-\textit{th} epoch, following the cosine function. For the detector, a similar training method was employed. Detector's weights were initialized with parameters trained on the COCO dataset~\cite{lin2014microsoft}, a large dataset with common objects marked by bounding boxes. The optimization objective incorporates both regression loss for box categories by weighted summation. A batch size of 8 was used due to larger memory consumption compared to the classifier and the total training schedule was 60 epochs.

\subsection*{Clinical evaluation results at the HMUH}

We provide in this section experimental results of the clinical evaluation at the HMUH. Table~\ref{tab:HMUH_agreement_ai} shows the inter-rater agreement among radiologists at the HMUH without the VinDr-CXR assistance. Table~\ref{tab:HMUH_agreement_with_ai} shows the inter-rater agreement among radiologists at the HMUH with the VinDr-CXR assistance. Table~\ref{tab:AI_rad_HMU_agreement_no_support} and Table~\ref{tab:AI_rad_HMU_agreement_support} report the agreement between the VinDr-CXR system and HMUH's radiologists without and with assistance, respectively.
 
\begin{table*}
    \centering
    \scriptsize{
    \caption{Inter-rater agreement among radiologists at the HMUH without the VinDr-CXR assistance.}
    \label{tab:HMUH_agreement_ai}
    \setlength{\tabcolsep}{2pt}
    \begin{tabular}{
    >{}p{70pt}|
    >{}p{25pt}|
    >{}p{70pt}|
    >{}p{25pt}|
    >{}p{70pt}|
    >{}p{25pt}|
    >{}p{70pt}|
    >{}p{70pt}}
    \hline
     & 
    \multicolumn{2}{l|}{ \hspace*{0.5cm} \textbf{Rater 1 \textit{vs}. Rater 2}} & 
    \multicolumn{2}{l|}{ \hspace*{0.5cm} \textbf{Rater 1 \textit{vs}. Rater 3}} & 
    \multicolumn{2}{l|}{ \hspace*{0.5cm} \textbf{Rater 2 \textit{vs}. Rater 3}} & 
     \\ \cline{2-7}
    \multirow{-2}{*}{ \textbf{Findings}} & 
    Agree & \hspace*{0.4cm} Cohen's $\kappa$ & 
   Agree & \hspace*{0.4cm} Cohen's $\kappa$ & 
   Agree & \hspace*{0.4cm} Cohen's $\kappa$ &  
    \multirow{-2}{*}{ \hspace*{0.6cm} Fleiss' $\kappa$}\\
    \hline
    Lung Tumor         & 0.955 & 0.617 (0.388, 0.846)  & 0.945 & 0.591 (0.372, 0.810)    & 0.960 & 0.646 (0.419, 0.873)    & 0.616 (0.536, 0.696)  \\
    Pneumonia          & 0.865 & 0.536 (0.388, 0.684)  & 0.815 & 0.312 (0.157, 0.468)    & 0.870 & 0.281 (0.080, 0.482)    & 0.383 (0.303, 0.463)  \\
    Tuberculosis       & 0.925 & 0.675 (0.522, 0.827)  & 0.915 & 0.385 (0.164, 0.605)    & 0.880 & 0.298 (0.111, 0.486)    & 0.474 (0.394, 0.554)  \\
    Other Diseases      & 0.775 & 0.470 (0.344, 0.596)  & 0.660 & 0.234 (0.105, 0.363)    & 0.735 & 0.440 (0.312, 0.567)    & 0.373 (0.293, 0.453)  \\
    No Finding         & 0.880 & 0.758 (0.668, 0.848)  & 0.865 & 0.730 (0.636, 0.825)    & 0.905 & 0.811 (0.731, 0.890)    & 0.766 (0.686, 0.846)  \\
    Aortic Enlargement & 0.955 & 0.549 (0.289, 0.810)  & 0.915 & -0.032 (-0.056, -0.007) & 0.940 & -0.027(-0.047, -0.008) & 0.207 (0.127, 0.287)  \\
    Atelectasis        & 0.980 & 0.490 (0.060, 0.920)  & 0.975 & -0.008 (-0.021, 0.005)  & 0.975 & -0.008 (-0.021, 0.005)  & 0.210 (0.130, 0.290)  \\
    Calcification      & 0.925 & 0.610 (0.432, 0.789)  & 0.920 & 0.556 (0.363, 0.749)    & 0.945 & 0.735 (0.586, 0.884)    & 0.638 (0.558, 0.718)  \\
    Cardiomegaly       & 0.890 & 0.277 (0.076, 0.478)  & 0.925 & 0.318 (0.067, 0.568)    & 0.885 & 0.404 (0.207, 0.601)    & 0.333 (0.253, 0.413)  \\
    Consolidation      & 0.935 & 0.216 (-0.046, 0.478) & 0.955 & 0.164 (-0.144, 0.471)   & 0.930 & 0.330 (0.066, 0.593)    & 0.249 (0.169, 0.329)  \\
    ILD                & 0.920 & 0.488 (0.275, 0.702)  & 0.815 & 0.131 (-0.028, 0.290)   & 0.845 & 0.352 (0.175, 0.529)    & 0.303 (0.223, 0.383)  \\
    Infiltration       & 0.885 & 0.000 (0.000, 0.000)  & 0.980 & 0.000 (0.000, 0.000)    & 0.885 & 0.118 (-0.056, 0.292)   & 0.030 (-0.050, 0.110) \\
    Lung Opacity       & 0.785 & 0.229 (0.072, 0.386)  & 0.785 & 0.103 (-0.020, 0.225)   & 0.870 & 0.020 (-0.114, 0.154)   & 0.116 (0.036, 0.196)  \\
    Nodule/Mass        & 0.955 & 0.688 (0.498, 0.877)  & 0.915 & 0.475 (0.267, 0.682)    & 0.900 & 0.490 (0.298, 0.682)    & 0.540 (0.460, 0.620)  \\
    Pleural Effusion   & 0.925 & 0.509 (0.298, 0.721)  & 0.945 & 0.674 (0.494, 0.853)    & 0.930 & 0.425 (0.178, 0.673)    & 0.547 (0.467, 0.627)  \\
    Pleural Thickening & 0.940 & 0.377 (0.106, 0.648)  & 0.975 & 0.275 (-0.164, 0.714)   & 0.925 & 0.102 (-0.100, 0.304)   & 0.245 (0.165, 0.325)  \\
    Pneumothorax       & 1.000 & 1.000 (1.000, 1.000)  & 1.000 & 1.000 (1.000, 1.000)    & 1.000 & 1.000 (1.000, 1.000)    & 1.000 (1.000, 1.000)  \\
    Pulmonary Fibrosis & 0.905 & 0.545 (0.367, 0.724)  & 0.895 & 0.343 (0.125, 0.560)    & 0.870 & 0.367 (0.177, 0.557)    & 0.423 (0.343, 0.503)  \\
    Other Lesions       & 0.890 & 0.175 (-0.024, 0.375) & 0.920 & 0.165 (-0.073, 0.402)   & 0.890 & 0.332 (0.120, 0.544)    & 0.232 (0.152, 0.312)  \\
    \hline
    \textbf{Mean}               & \textbf{0.910} & \textbf{0.485 (0.301, 0.669)}  & \textbf{0.901} & \textbf{0.338 (0.171, 0.504)}    & \textbf{0.902} & \textbf{0.374 (0.217, 0.532)}    & \textbf{0.404 (0.329, 0.480)} \\
    \hline
    \end{tabular}}
\end{table*}

\begin{table*}
    \centering
    \scriptsize{
    \caption{Inter-rater agreement among radiologists at the HMUH with the VinDr-CXR assistance.}
    \label{tab:HMUH_agreement_with_ai}
    \setlength{\tabcolsep}{2pt}
    \begin{tabular}{
    >{}p{70pt}|
    >{}p{25pt}|
    >{}p{70pt}|
    >{}p{25pt}|
    >{}p{70pt}|
    >{}p{25pt}|
    >{}p{70pt}|
    >{}p{70pt}}
    \hline
     & 
    \multicolumn{2}{l|}{ \hspace*{0.5cm} \textbf{Rater 1 \textit{vs}. Rater 2}} & 
    \multicolumn{2}{l|}{ \hspace*{0.5cm} \textbf{Rater 1 \textit{vs}. Rater 3}} & 
    \multicolumn{2}{l|}{ \hspace*{0.5cm} \textbf{Rater 2 \textit{vs}. Rater 3}} & 
     \\ \cline{2-7}
    \multirow{-2}{*}{ \textbf{Findings}} & 
    Agree & \hspace*{0.4cm} Cohen's $\kappa$ & 
    Agree & \hspace*{0.4cm} Cohen's $\kappa$ & 
    Agree & \hspace*{0.4cm} Cohen's $\kappa$ &  
    \multirow{-2}{*}{ \hspace*{0.6cm} Fleiss' $\kappa$}\\
    \hline
    Lung Tumor         & 0.955 & 0.617 (0.388, 0.846)  & 0.950 & 0.640 (0.433, 0.846)   & 0.965 & 0.702 (0.494, 0.910)   & 0.652 (0.572, 0.732)  \\
    Pneumonia          & 0.865 & 0.536 (0.388, 0.684)  & 0.820 & 0.338 (0.181, 0.494)   & 0.875 & 0.323 (0.121, 0.524)   & 0.403 (0.323, 0.483)  \\
    Tuberculosis       & 0.925 & 0.685 (0.537, 0.833)  & 0.915 & 0.385 (0.164, 0.605)   & 0.870 & 0.279 (0.100, 0.459)   & 0.471 (0.391, 0.551)  \\
    Other Diseases      & 0.775 & 0.470 (0.344, 0.596)  & 0.665 & 0.241 (0.111, 0.371)   & 0.740 & 0.449 (0.322, 0.576)   & 0.379 (0.299, 0.459)  \\
    No Finding         & 0.880 & 0.758 (0.668, 0.848)  & 0.870 & 0.740 (0.647, 0.833)   & 0.900 & 0.801 (0.719, 0.882)   & 0.766 (0.686, 0.846)  \\
    Aortic Enlargement & 0.945 & 0.565 (0.337, 0.793)  & 0.920 & 0.243 (0.001, 0.486)   & 0.935 & 0.103 (-0.134, 0.341)  & 0.340 (0.260, 0.420)  \\
    Atelectasis        & 0.975 & 0.432 (0.021, 0.842)  & 0.970 & -0.008 (-0.022, 0.005) & 0.975 & -0.008 (-0.021, 0.005) & 0.186 (0.106, 0.266)  \\
    Calcification      & 0.930 & 0.656 (0.490, 0.823)  & 0.940 & 0.694 (0.531, 0.856)   & 0.940 & 0.716 (0.564, 0.868)   & 0.689 (0.609, 0.769)  \\
    Cardiomegaly       & 0.885 & 0.267 (0.071, 0.463)  & 0.930 & 0.335 (0.076, 0.593)   & 0.885 & 0.405 (0.210, 0.601)   & 0.333 (0.253, 0.413)  \\
    Consolidation      & 0.935 & 0.216 (-0.046, 0.478) & 0.955 & 0.164 (-0.144, 0.471)  & 0.930 & 0.330 (0.066, 0.593)   & 0.249 (0.169, 0.329)  \\
    ILD                & 0.930 & 0.575 (0.376, 0.773)  & 0.815 & 0.161 (-0.002, 0.324)  & 0.855 & 0.412 (0.238, 0.585)   & 0.361 (0.281, 0.441)  \\
    Infiltration       & 0.880 & 0.000 (0.000, 0.000)  & 0.980 & 0.000 (0.000, 0.000)   & 0.880 & 0.112 (-0.055, 0.280)  & 0.026 (-0.054, 0.106) \\
    Lung Opacity       & 0.775 & 0.212 (0.057, 0.367)  & 0.780 & 0.099 (-0.020, 0.219)  & 0.865 & 0.017 (-0.113, 0.146)  & 0.106 (0.026, 0.186)  \\
    Nodule/Mass        & 0.950 & 0.662 (0.469, 0.855)  & 0.915 & 0.500 (0.299, 0.701)   & 0.905 & 0.525 (0.338, 0.712)   & 0.554 (0.474, 0.634)  \\
    Pleural Effusion   & 0.925 & 0.509 (0.298, 0.721)  & 0.945 & 0.674 (0.494, 0.853)   & 0.930 & 0.425 (0.178, 0.673)   & 0.547 (0.467, 0.627)  \\
    Pleural Thickening & 0.930 & 0.338 (0.084, 0.593)  & 0.975 & 0.275 (-0.164, 0.714)  & 0.915 & 0.089 (-0.092, 0.270)  & 0.219 (0.139, 0.299)  \\
    Pneumothorax       & 1.000 & 1.000 (1.000, 1.000)  & 1.000 & 1.000 (1.000, 1.000)   & 1.000 & 1.000 (1.000, 1.000)   & 1.000 (1.000, 1.000)  \\
    Pulmonary fibrosis & 0.890 & 0.484 (0.301, 0.666)  & 0.895 & 0.343 (0.125, 0.560)   & 0.875 & 0.403 (0.216, 0.590)   & 0.413 (0.333, 0.493)  \\
    Other Lesions       & 0.895 & 0.189 (-0.013, 0.392) & 0.915 & 0.158 (-0.066, 0.383)  & 0.890 & 0.362 (0.153, 0.571)   & 0.248 (0.168, 0.328)  \\
    \hline
    \textbf{Mean}  & \textbf{0.908} & \textbf{0.483 (0.304, 0.662)}  & \textbf{0.903} & \textbf{0.367 (0.192, 0.543)}   & \textbf{0.902} & \textbf{0.392 (0.227, 0.557)}   & \textbf{0.418 (0.342, 0.494)} \\
    \hline
    \end{tabular}}
\end{table*}

\begin{table*}
    \centering
    \scriptsize{
    \caption{Agreement between the VinDr-CXR system and HMUH's radiologists without assistance.}
    \label{tab:AI_rad_HMU_agreement_no_support}
    \setlength{\tabcolsep}{2pt}
    \begin{tabular}{
    >{}p{70pt}|
    >{}p{30pt}|
    >{}p{70pt}|
    >{}p{30pt}|
    >{}p{70pt}|
    >{}p{30pt}|
    >{}p{70pt}}
    \hline
      & 
    \multicolumn{2}{l|}{\hspace*{0.5cm} \textbf{Rater 1 \textit{vs}. AI}} & 
    \multicolumn{2}{l|}{ \hspace*{0.5cm} \textbf{Rater 2 \textit{vs}. AI}} & 
    \multicolumn{2}{l}{ \hspace*{0.5cm} \textbf{Rater 3 \textit{vs}. AI}}  \\ \cline{2-7}
    \multirow{-2}{*}{\textbf{Findings}} & 
    Agree & \hspace*{0.4cm} Cohen's $\kappa$ & 
    Agree & \hspace*{0.4cm} Cohen's $\kappa$ & 
    Agree & \hspace*{0.4cm} Cohen's $\kappa$  \\
    \hline
    Lung Tumor         & 0.940 & 0.427 (0.164, 0.690)    & 0.975 & 0.693 (0.439, 0.947)    & 0.945 & 0.451 (0.182, 0.719)  \\
    Pneumonia          & 0.845 & 0.513 (0.366, 0.661)    & 0.890 & 0.559 (0.396, 0.721)    & 0.870 & 0.417 (0.239, 0.594)  \\
    Tuberculosis       & 0.910 & 0.628 (0.472, 0.784)    & 0.945 & 0.793 (0.675, 0.910)    & 0.865 & 0.271 (0.096, 0.446)  \\
    Other Diseases      & 0.765 & 0.390 (0.248, 0.532)    & 0.840 & 0.640 (0.528, 0.752)    & 0.685 & 0.314 (0.181, 0.447)  \\
    No finding         & 0.860 & 0.721 (0.625, 0.816)    & 0.870 & 0.742 (0.652, 0.833)    & 0.845 & 0.689 (0.589, 0.790)  \\
    Aortic Enlargement & 0.945 & 0.616 (0.409, 0.823)    & 0.930 & 0.430 (0.188, 0.672)    & 0.910 & 0.154 (-0.058, 0.366) \\
    Atelectasis        & 0.970 & 0.235 (-0.167, 0.636)   & 0.980 & 0.490 (0.060, 0.920)    & 0.985 & 0.395 (-0.146, 0.936) \\
    Calcification      & 0.900 & 0.237 (0.011, 0.462)    & 0.895 & 0.354 (0.146, 0.562)    & 0.920 & 0.463 (0.246, 0.680)  \\
    Cardiomegaly       & 0.875 & 0.203 (0.017, 0.389)    & 0.925 & 0.674 (0.520, 0.828)    & 0.870 & 0.340 (0.146, 0.535)  \\
    Consolidation      & 0.960 & 0.318 (-0.026, 0.662)   & 0.925 & 0.310 (0.052, 0.568)    & 0.935 & 0.201 (-0.073, 0.476) \\
    ILD                & 0.850 & 0.178 (-0.006, 0.361)   & 0.880 & 0.433 (0.245, 0.621)    & 0.815 & 0.286 (0.113, 0.459)  \\
    Infiltration       & 0.885 & 0.000 (0.000, 0.000)    & 0.880 & 0.410 (0.217, 0.604)    & 0.885 & 0.118 (-0.056, 0.292) \\
    Lung Opacity       & 0.785 & 0.179 (0.030, 0.329)    & 0.880 & 0.269 (0.060, 0.479)    & 0.900 & 0.045 (-0.124, 0.215) \\
    Nodule/Mass        & 0.880 & 0.378 (0.190, 0.566)    & 0.875 & 0.442 (0.261, 0.623)    & 0.855 & 0.390 (0.211, 0.570)  \\
    Pleural Effusion   & 0.945 & 0.640 (0.447, 0.833)    & 0.940 & 0.423 (0.154, 0.691)    & 0.970 & 0.754 (0.565, 0.943)  \\
    Pleural Thickening & 0.945 & -0.028 (-0.045, -0.011) & 0.895 & -0.045 (-0.072, -0.018) & 0.970 & 0.239 (-0.158, 0.635) \\
    Pneumothorax       & 0.990 & 0.496 (-0.104, 1.000)   & 0.990 & 0.496 (-0.104, 1.000)   & 0.990 & 0.496 (-0.104, 1.000) \\
    Pulmonary Fibrosis & 0.880 & 0.302 (0.093, 0.511)    & 0.865 & 0.375 (0.188, 0.562)    & 0.865 & 0.196 (-0.003, 0.396) \\
    Other Lesions       & 0.935 & -0.033 (-0.052, -0.015) & 0.895 & 0.235 (0.023, 0.447)    & 0.905 & 0.051 (-0.126, 0.228) \\
    \hline
    \textbf{Mean}               & \textbf{0.898} & \textbf{0.337 (0.141, 0.528) }   & \textbf{0.909} & \textbf{0.459 (0.244, 0.670) }   & \textbf{0.894} & \textbf{0.330 (0.090, 0.565)} \\
    \hline
    \end{tabular}}
\end{table*}

\begin{table*}
    \centering
    \scriptsize{
    \caption{Agreement between the VinDr-CXR system and HMUH's radiologists with assistance.}
    \label{tab:AI_rad_HMU_agreement_support}
    \setlength{\tabcolsep}{2pt}
    \begin{tabular}{
    >{}p{70pt}|
    >{}p{30pt}|
    >{}p{70pt}|
    >{}p{30pt}|
    >{}p{70pt}|
    >{}p{30pt}|
    >{}p{70pt}}
    \hline
     & 
    \multicolumn{2}{l|}{ \hspace*{0.5cm} Rater 1 \textit{vs}. AI} & 
    \multicolumn{2}{l|}{ \hspace*{0.5cm} Rater 2 \textit{vs}. AI} & 
    \multicolumn{2}{l }{ \hspace*{0.5cm} Rater 3 \textit{vs}. AI} \\ \cline{2-7}
    \multirow{-2}{*}{Findings} &
    Agree & \hspace*{0.4cm} Cohen's $\kappa$ & 
    Agree & \hspace*{0.4cm} Cohen's $\kappa$ & 
    Agree & \hspace*{0.4cm} Cohen's $\kappa$ \\
    \hline
    Lung Tumor         & 0.940 & 0.427 (0.164, 0.690)    & 0.975 & 0.693 (0.439, 0.947)  & 0.950 & 0.523 (0.267, 0.778)  \\
    Pneumonia          & 0.845 & 0.513 (0.366, 0.661)    & 0.890 & 0.559 (0.396, 0.721)  & 0.875 & 0.447 (0.271, 0.623)  \\
    Tuberculosis       & 0.910 & 0.628 (0.472, 0.784)    & 0.945 & 0.798 (0.683, 0.913)  & 0.865 & 0.271 (0.096, 0.446)  \\
    Other Diseases      & 0.765 & 0.390 (0.248, 0.532)    & 0.840 & 0.640 (0.528, 0.752)  & 0.690 & 0.322 (0.188, 0.455)  \\
    No Finding         & 0.860 & 0.721 (0.625, 0.816)    & 0.870 & 0.742 (0.652, 0.833)  & 0.850 & 0.700 (0.601, 0.799)  \\
    Aortic Enlargement & 0.955 & 0.718 (0.544, 0.893)    & 0.940 & 0.542 (0.315, 0.769)  & 0.915 & 0.231 (-0.003, 0.464) \\
    Atelectasis        & 0.965 & 0.205 (-0.162, 0.571)   & 0.980 & 0.490 (0.060, 0.920)  & 0.985 & 0.395 (-0.146, 0.936) \\
    Calcification      & 0.905 & 0.343 (0.119, 0.566)    & 0.895 & 0.354 (0.146, 0.562)  & 0.925 & 0.511 (0.302, 0.721)  \\
    Cardiomegaly       & 0.875 & 0.203 (0.017, 0.389)    & 0.940 & 0.743 (0.605, 0.881)  & 0.875 & 0.354 (0.158, 0.550)  \\
    Consolidation      & 0.960 & 0.318 (-0.026, 0.662)   & 0.925 & 0.310 (0.052, 0.568)  & 0.935 & 0.201 (-0.073, 0.476) \\
    ILD                & 0.865 & 0.275 (0.080, 0.469)    & 0.885 & 0.466 (0.281, 0.651)  & 0.820 & 0.315 (0.143, 0.487)  \\
    Infiltration       & 0.885 & 0.000 (0.000, 0.000)    & 0.885 & 0.446 (0.255, 0.636)  & 0.885 & 0.118 (-0.056, 0.292) \\
    Lung Opacity       & 0.780 & 0.174 (0.027, 0.321)    & 0.885 & 0.318 (0.108, 0.527)  & 0.900 & 0.045 (-0.124, 0.215) \\
    Nodule/Mass        & 0.875 & 0.364 (0.176, 0.551)    & 0.875 & 0.442 (0.261, 0.623)  & 0.870 & 0.461 (0.286, 0.636)  \\
    Pleural Effusion   & 0.945 & 0.640 (0.447, 0.833)    & 0.940 & 0.423 (0.154, 0.691)  & 0.970 & 0.754 (0.565, 0.943)  \\
    Pleural Thickening & 0.945 & -0.028 (-0.045, -0.011) & 0.895 & 0.045 (-0.116, 0.205) & 0.970 & 0.239 (-0.158, 0.635) \\
    Pneumothorax       & 0.990 & 0.496 (-0.104, 1.000)   & 0.990 & 0.496 (-0.104, 1.000) & 0.990 & 0.496 (-0.104, 1.000) \\
    Pulmonary Fibrosis & 0.880 & 0.302 (0.093, 0.511)    & 0.860 & 0.364 (0.178, 0.549)  & 0.865 & 0.196 (-0.003, 0.396) \\
    Other Lesions       & 0.940 & -0.030 (-0.048, -0.012) & 0.895 & 0.235 (0.023, 0.447)  & 0.915 & 0.223 (-0.017, 0.463) \\
    \hline
    \textbf{Mean}              & \textbf{0.899} & \textbf{0.350 (0.158, 0.538)}    & \textbf{0.911} & \textbf{0.479 (0.259, 0.694)}  & \textbf{0.897} & \textbf{0.358 (0.115, 0.595)} \\
    \hline
    \end{tabular}}
\end{table*}

\subsection*{Examples of CXR scans from internal and external datasets}

\begin{figure*}
\centering
\includegraphics[width=15cm,height=3cm]{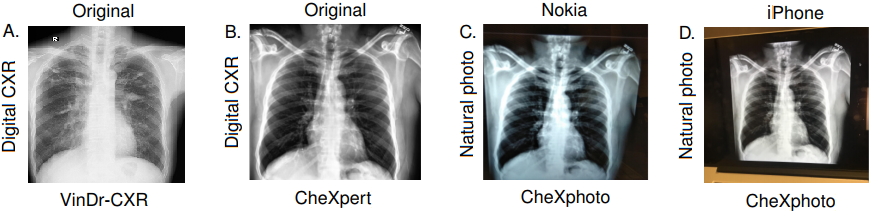}
\caption{Examples of CXR scans from internal and external datasets. (A) An original CXR scan in the DICOM format from the VinDr-CXR dataset. (B) An original CXR scan in Portable Graphics Format (.PNG) format from CheXpert~\cite{irvin2019chexpert} dataset. (C) A CXR image produced by a Nokia phone from CheXphoto~\cite{phillips2020chexphoto} dataset. (D) A CXR image produced by iPhone from CheXphoto~\cite{phillips2020chexphoto} dataset.}
\label{fig:external_CXR}
\end{figure*}

Figure~\ref{fig:external_CXR} shows several examples of CXR scans from internal (VinDr-CXR) and external datasets (CheXpert~\cite{irvin2019chexpert} and CheXphoto~\cite{phillips2020chexphoto}).

\subsection*{Representative cases from the reader study}

\begin{figure*}
\centering
\includegraphics[width=12cm,height=13cm]{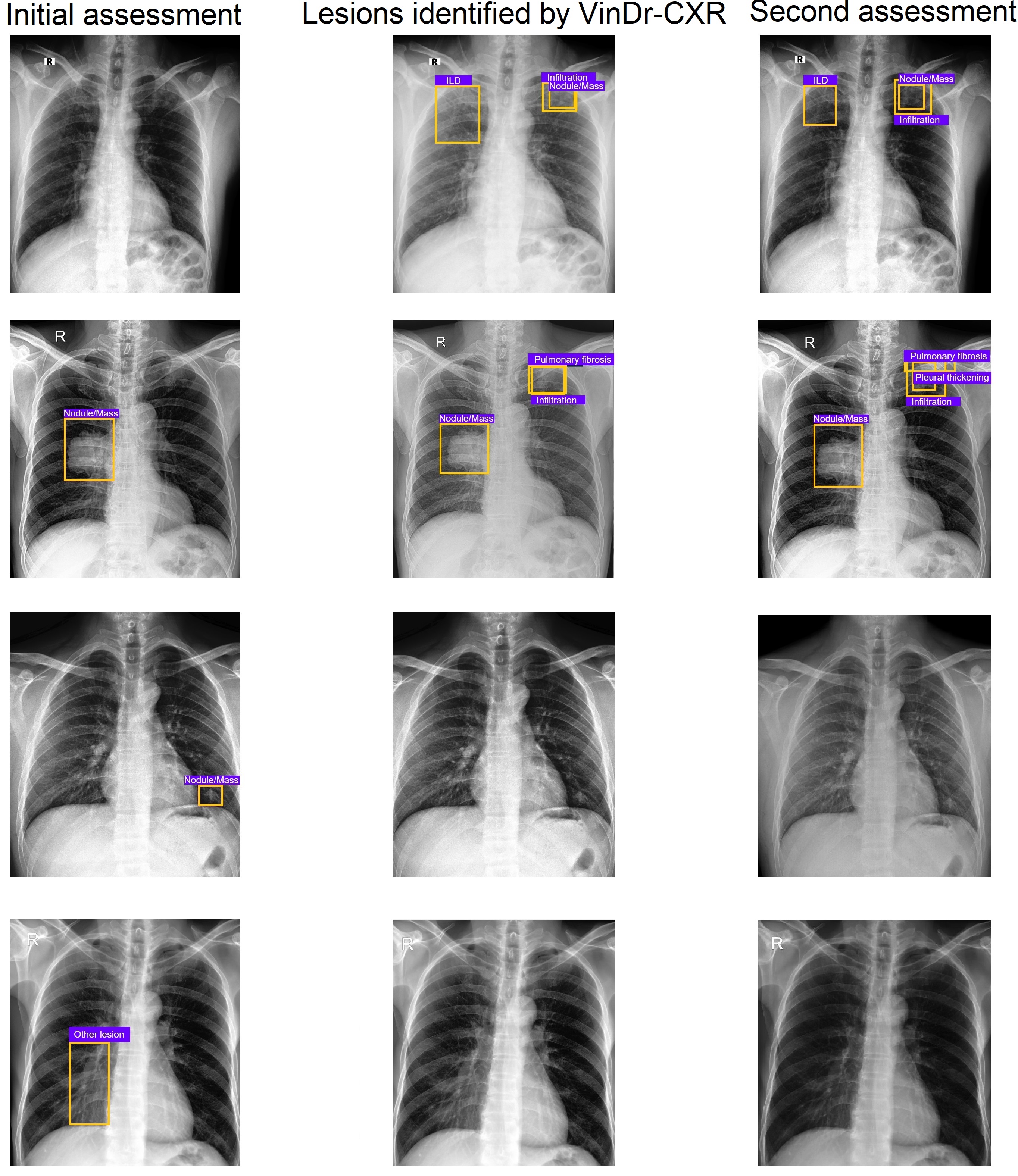}
\caption{Representative cases from our reader study. The first column shows CXRs and lesions marked by participating radiologists for the first assessment. The middle column shows lesions identified by the VinDr-CXR system. The last column shows the final decision of the radiologists after consulting the VinDr-CXR's result. In many cases, radiologists have changed their previous decisions by adding or removing lesions.}
\label{fig:reader_study_vis}
\end{figure*}

Several representative cases from our reader study are provided in Figure~\ref{fig:reader_study_vis}. In many cases, radiologists have changed their previous decisions by adding or removing lesions.

\subsection*{VinDr-CXR interface}

\begin{figure*}
\centering
\includegraphics[width=17cm,height=7cm]{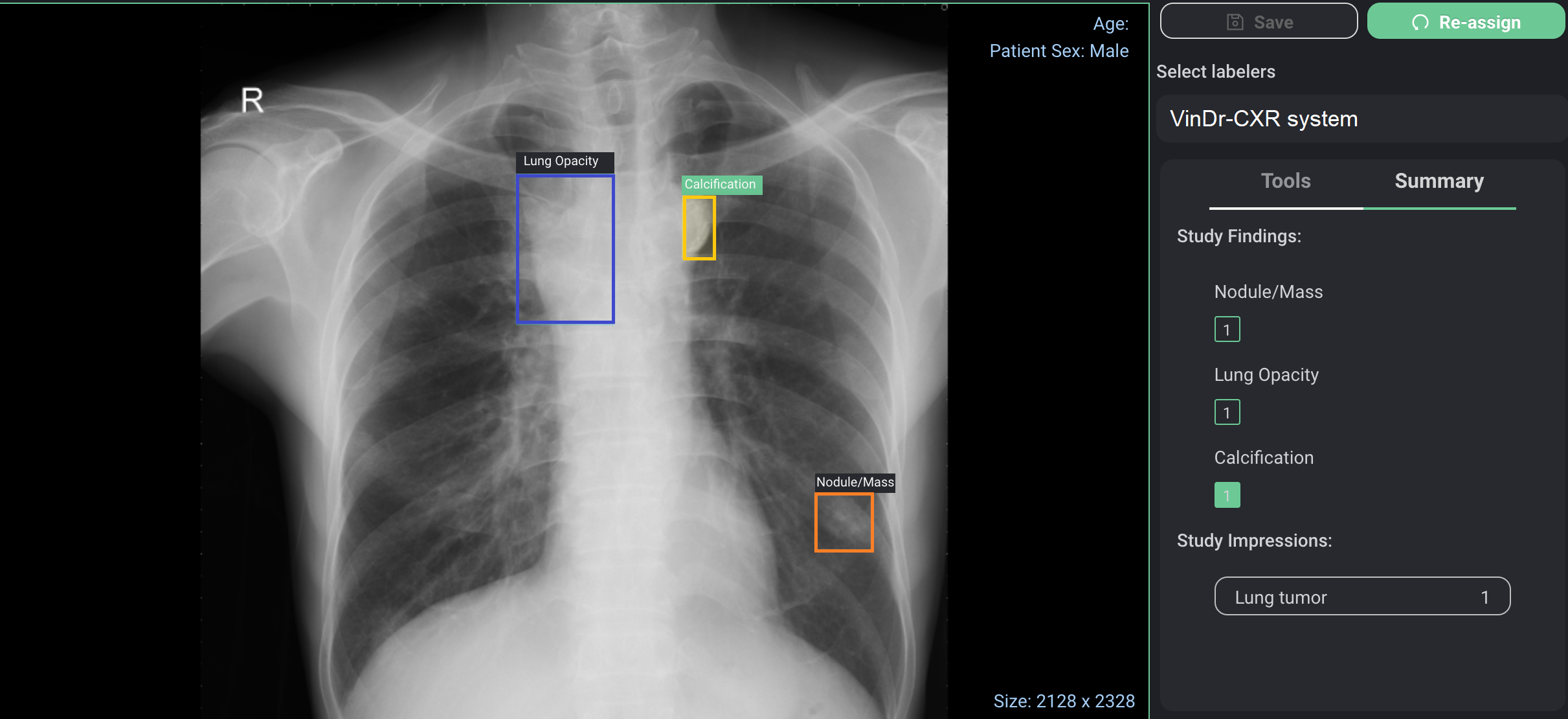}
\caption{VinDr-CXR interface. The clinical findings detected by the VinDr-CXR system are listed on the interface, and bounding box predictions are displayed on the image.}
\label{fig:vindr-cxr-interface}
\end{figure*}

Figure~\ref{fig:vindr-cxr-interface} shows VinDr-CXR interface that allows displayed detected findings on the image via bounding box predictions.  

\subsection*{Data availability}
\label{sec:data_availability}

To facilitate a wide range of research topics in computer vision and medical imaging, we made the VinDr-CXR dataset (18,000 studies) publicly available through through PhysioNet at \url{https://physionet.org/content/vindr-cxr/1.0.0/}. The image and annotation quality of the dataset can be visually check via our project webpage at \url{https://vindr.ai/datasets/cxr}. The CheXpert dataset is publicly available at \url{https://stanfordmlgroup.github.io/competitions/chexpert/}. The CheXphoto dataset is publicly available at \url{https://stanfordmlgroup.github.io/competitions/chexphoto/}.

\subsection*{Code availability}

Implementation of our work is based on the following open source repositories: Pytorch: \url{https://pytorch.org}; OpenCV \url{https://opencv.org/}; Pydicom: \url{https://pydicom.github.io/}. The source code used to train the VinDr-CXR system is a part of a commercial software product and not available to the public. The commercial version of VinDr-CXR can be freely tried through an online demonstration at \url{https://vindr.ai/}. All performance metrics were calculated with the support of scikit-learn \url{https://scikit-learn.org/}. Our labeling framework called VinDr Lab was made as open-source software and available for downloading at \url{https://vindr.ai/vindr-lab}.

\section*{Acknowledgment}

This research was supported by Vingroup Big Data Institute. The authors would like to acknowledge the Hanoi Medical University Hospital (HMUH) and the Hospital 108 (H108) for providing us access to their image databases and for agreeing to make the VinDr-CXR dataset publicly available. In particular, we would like to thank all of our radiologists, physicians, and technicians, who participated in this project.

\newpage 
\bibliography{references.bib}

\begin{IEEEbiography}[{\includegraphics[width=1in,height=1.25in,clip,keepaspectratio]{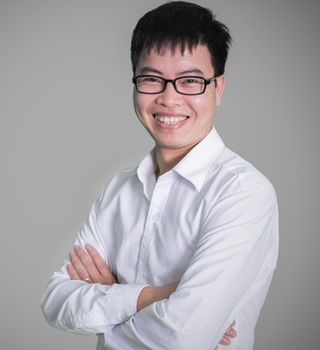}}]{Hieu H. Pham} Dr. Pham Huy Hieu is a Teaching Fellow at the College of Engineering and Computer Science (CECS), VinUniversity, and serves as Associate Director at VinUni-Illinois Smart Health Center. He received his Ph.D. in Computer Science from the Toulouse Computer Science Research Institute (IRIT), University of Toulouse, France, in 2019. Previously, he earned the Degree of Engineer in Industrial Informatics from Hanoi University of Science and Technology (HUST), Vietnam, in 2016. His research interests include Computer Vision, Machine Learning, Medical Image Analysis, and their applications in Smart Healthcare. He is the author, co-author of 30 scientific articles appeared in about 20 conferences and journals such as Computer Vision and Image Understanding, Neurocomputing, International Conference on Medical Image Computing and Computer-Assisted Intervention (MICCAI), Medical Imaging with Deep Learning (MIDL), IEEE International Conference on Image Processing (ICIP), and IEEE International Conference on Computer Vision (ICCV). He is also currently serving as Reviewers for MICCAI, ICCV, CVPR, IET Computer Vision Journal (IET-CVI), IEEE Journal of Biomedical and Health Informatics (JBHI), and Nature Scientific Reports. Before joining VinUniversity, Dr. Hieu worked at Vingroup Big Data Institute (VinBigData) as a Research Scientist and Head of the Fundamental Research Team. With this position, he led several research projects on Medical AI, including collecting various types of medical data, managing and annotating data, and developing new AI solutions for medical analysis.
\end{IEEEbiography}

\begin{IEEEbiography}[{\includegraphics[width=1in,height=1.25in,clip,keepaspectratio]{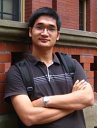}}]{Ha Q. Nguyen} Dr. Ha Q. Nguyen was born in Hai Phong, Vietnam, in 1983. He received the B.S. degree in mathematics from the Hanoi National University of Education, Hanoi, Vietnam, the S.M. degree in electrical engineering and computer science from the Massachusetts Institute of Technology, Cambridge, MA, USA, and the Ph.D. degree in electrical and computer engineering from the University of Illinois at Urbana-Champaign, Champaign, IL, USA, in 2005, 2009, and 2014, respectively. During 2009–2011, he was a Lecturer in electrical engineering with the International University, Vietnam National University, Ho Chi Minh City, Vietnam, during 2014–2017, he was a Postdoctoral Research Associate with the Biomedical Imaging Group, Ecole Polytechnique Federale de Lausanne, Lausanne, Switzerland, and during 2017–2018, he was a Signal Processing Engineer with the Viettel Research and Development Institute, Hanoi, Vietnam. He is currently the Head of Medical Imaging Department at the Vingroup Big Data Institute, Hanoi, Vietnam. His research interests include medical image analysis, machine learning, computational imaging, and data compression. Dr. Nguyen was a Fellow of Vietnam Education Foundation, cohort 2007. He was the recipient of the Best Student Paper Award (second prize) of the IEEE International Conference on Acoustics, Speech and Signal Processing in 2014 for his paper (with P.A. Chou and Y. Chen) on the compression of human body sequences using graph wavelet filter banks.
\end{IEEEbiography}

\begin{IEEEbiography}[{\includegraphics[width=1in,height=1.25in,clip,keepaspectratio]{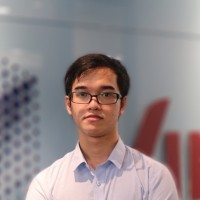}}]{Hieu T. Nguyen} Hieu T. Nguyen was pursuing the B.S. degree in Computer Science from the Department of Computer Science, Hanoi University of Science and Technology (HUST) Hanoi, Vietnam. He is an incoming CS PhD student at Northeastern University, USA. He is also a research intern at Medical Imaging Department at the Vingroup Big Data Institute (VinBDI), Hanoi, Vietnam. His research interests include medical image analysis and deep learning techniques.
\end{IEEEbiography}

\begin{IEEEbiography}[{\includegraphics[width=1in,height=1.25in,clip,keepaspectratio]{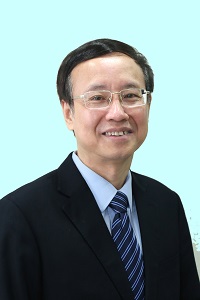}}]{Lam Khanh} Assoc. Prof. Dr. Lam Khanh -- Major General Lam Khanh, Deputy Director of the 108 Military Central Hospital. He received his Ph.D. in medical imaging from the National Institute for Basic Biology, Japan. His research focuses on MEG and Neuro MRI imaging. He is currently working on brain magnetic resonance imaging, interventional imaging, application of multi-sequence CT in diagnosis and treatment.
\end{IEEEbiography}

\begin{IEEEbiography}[{\includegraphics[width=1in,height=1.25in,clip,keepaspectratio]{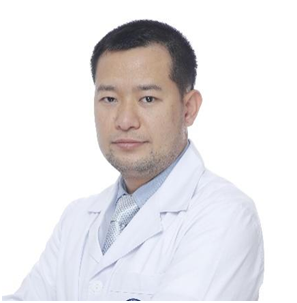}}]{Linh T. Le} Dr. Linh T. Le is currently the Director of Department of Radiology, Hanoi Medical University Hospital, Hanoi, Vietnam. He is also a faculty member at Hanoi Medical University Hospital. He is an experienced radiologist with 20 years working in medical imaging analysis (X-ray analysis, CT, MRI interpretation). He is currently working on chest X-ray imaging and many other imaging modalities.
\end{IEEEbiography}

\EOD
\end{document}